\documentclass[12pt]{article}

\usepackage{times}
\usepackage{epsf}
\usepackage{graphics}
\usepackage{graphicx}
\usepackage{xcolor}
\usepackage{caption} 
\usepackage{lineno}

\usepackage{pdflscape}
\usepackage{afterpage}
\usepackage{capt-of}

\usepackage[
  top    = 0.2in,
  bottom = 0.25in,
  left   = 1.in,
  right  = 1.in,
  headsep=5mm,
  includeheadfoot,
  ]{geometry}

\newcounter{lastnote}

\title{Role of the North Atlantic in Indian Monsoon Droughts}

\author   {P. J. Borah$^{1}$,   V.   Venugopal$^{1,2, \ast}$,
  J. Sukhatme$^{1}$, P. Muddebihal$^{1}$ 
  and B. N. Goswami$^{3}$\\\\
  \normalsize{$^{1}$Centre for Atmospheric and Oceanic Sciences \&}\\
 \normalsize{Divecha Centre for Climate Change}\\
\normalsize{Indian Institute of Science, Bangalore 560012, India.}\\\\
\normalsize{$^{2}$ Interdisciplinary Centre for Water Research}\\
\normalsize{Indian  Institute of Science,  Bangalore, 560012,  India.}\\\\
\normalsize{$^{3}$ Department of Physics, Cotton University, 
  Guwahati 781001, India.}\\\\
\normalsize{$^\ast$Correspondence:   E-mail: venu.vuruputur@gmail.com
  --or-- venu@iisc.ac.in}}

\date{}

\begin{document} 


\baselineskip 24pt


\maketitle 


\begin{abstract}
The forecast  of Indian  monsoon droughts has  been predicated  on the
notion of a  season-long rainfall deficit linked to  warm anomalies in
the equatorial Pacific.  Here, we show that in fact nearly half of all
droughts over the past century were sub-seasonal, and characterized by
an abrupt decline in late-season rainfall.  Furthermore, the potential
driver of  this class of  droughts is a  coherent cold anomaly  in the
North  Atlantic Ocean.   The  vorticity forcing  associated with  this
oceanic  marker extends  through  the depth  of  the troposphere,  and
results in a wavetrain which curves towards the equator and extends to
East-Asia.  This upper-level response  triggers an anomalous low-level
anticyclonic  circulation   late  in  the  season   over  India.  This
teleconnection  from the  midlatitudes offers  an avenue  for improved
predictability of monsoon droughts.
\end{abstract}

\clearpage

Embedded  within  the  interannual and  intraseasonal  variability  of
Indian         summer          monsoon         rainfall         (ISMR)
\cite{partha78,yasunari81,krishna2000,gadgil2003,hoyos2007}        are
complex space-time patterns pertaining  to its extreme states, namely,
floods and droughts \cite{gadgil2003,sikka99,goswami2005south}.  Given
that the socio-economic fabric of  one sixth of the world's population
is intricately  tied to  the state  of the  monsoon \cite{gadgil2006},
these  extremes, especially  droughts,  have a  devastating impact  on
agriculture and the economy.  Against  this backdrop, despite its high
potential predictability \cite{charney1981}, the  forecast of ISMR one
season      in     advance      remains     a      grand     challenge
\cite{rajeevan2012,ravi2013,wang2016}.  For instance, it is known that
anomalously warm waters in  the equatorial central/eastern Pacific (El
Ni\~{n}o)     are    associated     with     some    ISMR     droughts
\cite{rasmusson83,webster98,krishnakumar2006,fan2017}.   However,  the
dynamics, or what drives droughts  not associated with El Ni\~{n}o has
remained mostly unknown.  Added to  this is the prevailing notion that
ISMR droughts  are characterized  by large-scale  season-long rainfall
deficit.  Indeed, a deeper insight into droughts via their sub-seasonal
evolution is  essential not only  for a better understanding  of these
extremes,  but also  practically for  targeted improvement  of general
circulation  models  in  realizing  the potential  for  improved  ISMR
prediction.

\begin{figure}
  \begin{center}
\includegraphics[width= \textwidth]{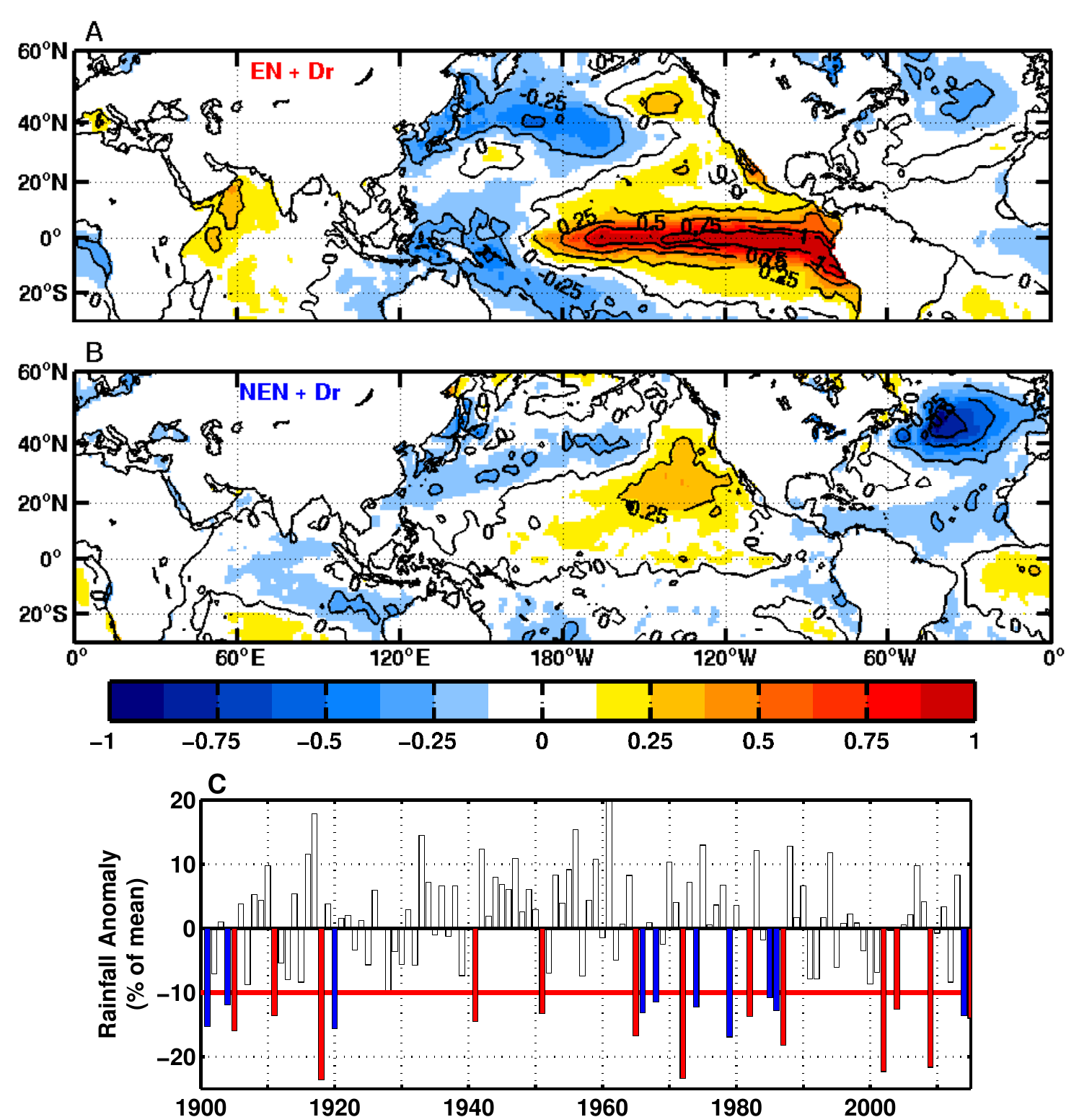}
    \caption{Spatial distribution  of anomalies of detrended  JJAS sea
      surface temperature  (SST) for the  two types of  Indian monsoon
      droughts: (A) EN + Dr; (B) NEN + Dr.  The detrending is based on
      removal of a  linear trend for 1901-2015  at each grid-location,
      and the  maps shown are  the average of fluctuations  around the
      trend,   for   the   years   listed   in   \textcolor{red}{Table
        \ref{tab:tab_s1}}.   Based  on  the 1$^{\circ}$,  monthly  SST
      product from the Hadley Centre \cite{hadsstref}.  Colorbar is in
      $^{\circ}$C.  (C) Interannual variation of anomalies of seasonal
      (JJAS)  monsoon rainfall  based on  the IITM  homogeneous Indian
      monthly  rainfall  dataset  \cite{iitmdataset}  for  the  period
      1901-2015.   The  horizontal  red  line  at  -10\%  denotes  the
      threshold   for   defining  a   drought,   as   per  the   India
      Meteorological Department  (IMD).  The departures marked  in red
      (blue) indicate droughts that occurred during an El Ni\~{n}o (no
      El Ni\~{n}o) year.}
\label{fig:fig_1}
  \end{center}
  \end{figure}

Over  the past  century, India  has  experienced 23  droughts and,  as
identified in  \textcolor{red}{Table \ref{tab:tab_s1}},  13 of these  occurred with,
and  10 that  occurred without,  an El  Ni\~{n}o \cite{varikoden2014}.
These two  flavors of droughts are  henceforth referred to as  EN + Dr
and NEN + Dr, respectively.   This sea surface temperature (SST) based
classification   is   confirmed   in   Figs.    \ref{fig:fig_1}a   and
\ref{fig:fig_1}b,  which  show the  mean  of  detrended SST  anomalies
during  June  through  September  (JJAS)   for  the  years  listed  in
\textcolor{red}{Table   \ref{tab:tab_s1}}.     Both   types   of    droughts   (Fig.
\ref{fig:fig_1}c), though  significantly different from  normal years,
are   statistically  indistinguishable   from   each   other  at   5\%
significance   level  ($p$-value   =   0.09)  on   a  seasonal   scale
(\textcolor{red}{Fig. \ref{fig:fig_s1}}).   Given the  disparate oceanic  conditions
associated  with  the two  kinds  of  droughts, we  investigate  their
sub-seasonal evolution to similar final seasonal states.

\begin{figure}
\begin{center}
\includegraphics[width= 0.75\textwidth]{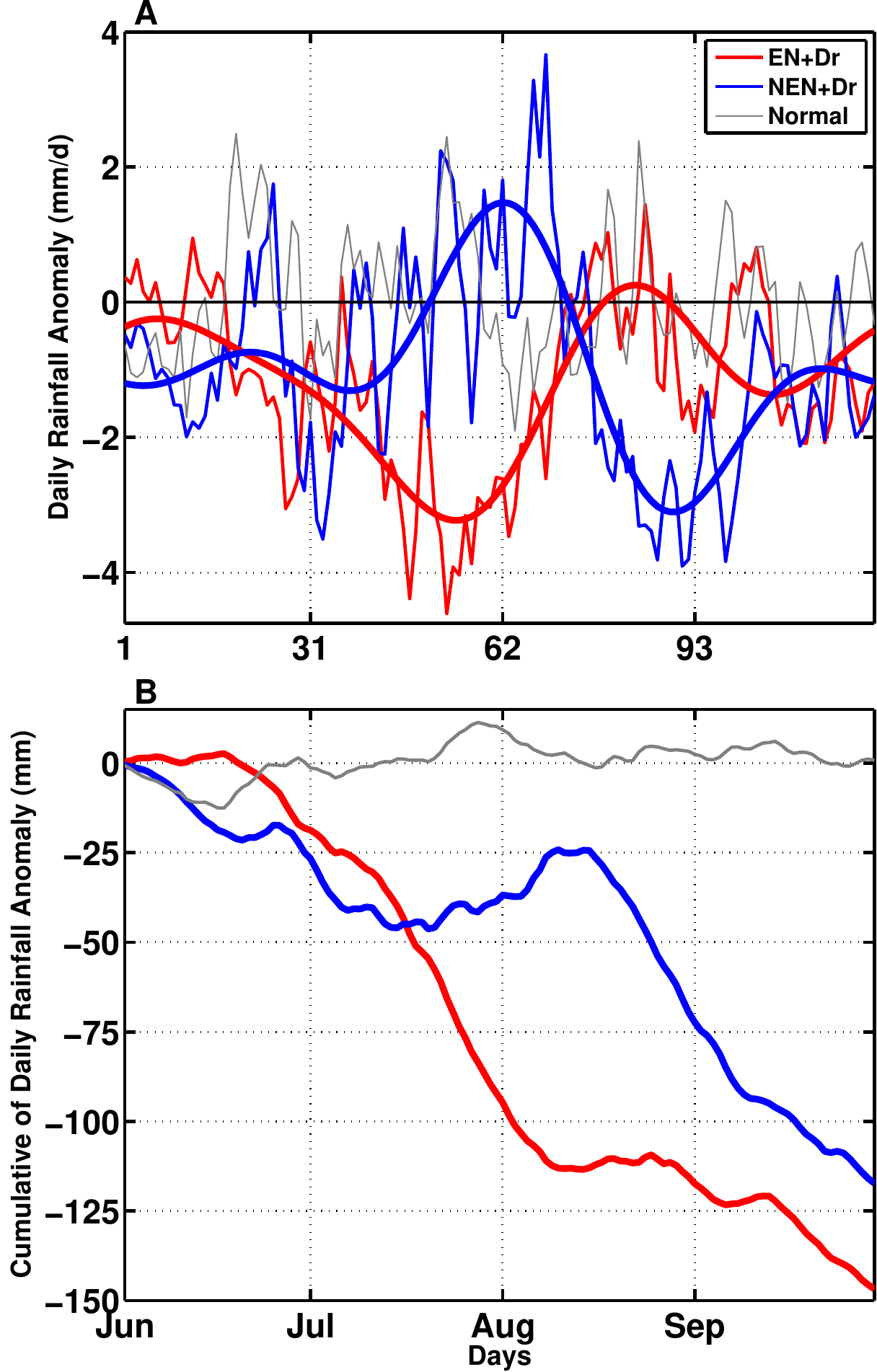}
  \caption{(A) Temporal evolution of  anomalies of daily area-averaged
    rainfall over central  India (16.5N-26.5N, 74.5E-86.5E; land-only)
    during JJAS, for the two  categories of droughts, along with their
    respective leading three harmonics ($\sim$ 120, 60, 40 days; thick
    red and blue).  The thin red  and blue curves shown are an average
    of daily anomalies of  13 EN + Dr and 10 NEN +  Dr years listed in
    \textcolor{red}{Table  S1}.   (B)  Cumulative  of  daily  rainfall
    anomalies shown in  (A).  The average of  daily rainfall anomalies
    for a normal year (based on 18 randomly chosen years with seasonal
    rainfall  anomaly between  -5\% and  5\%) and  its cumulative  are
    shown  as  a grey  line  in  both  panels.  Based on  IMD  gridded
    1-degree, daily rainfall data from 1901-2015.}
\label{fig:fig_2}
\end{center}
\end{figure}

 Figure  \ref{fig:fig_2}a   shows  the  temporal  behavior   of  daily
 anomalies  of  area-averaged  central  India  rainfall  (16.5N-26.5N;
 74.5E-86.5E; see  methods) during  JJAS, for  the two  categories, in
 comparison  to normal  years.  The  daily anomalies  for each  of the
 drought years  are shown in  \textcolor{red}{Fig. \ref{fig:fig_s2}},
 from  which   the  averages  shown  in   Fig.   \ref{fig:fig_2}a  are
 estimated.  The  difference in the  evolution of the two  droughts is
 immediately evident.  Specifically, in the EN + Dr category (thin red
 curve in Fig.  \ref{fig:fig_2}a),  negative rainfall anomalies, i.e.,
 departures from the daily climatological mean, set in during mid-June
 and persist till early August.  On the  other hand, in NEN + Dr (thin
 blue curve in Fig.  \ref{fig:fig_2}a),  anomalies have more rapid and
 intense fluctuations culminating in an abrupt drop (also, long break;
 \textcolor{black}{mid-to-late-August  period highlighted  by the  blue
   ellipse   in}   \textcolor{red}{Fig. \ref{fig:fig_s2}b})   during
 mid-August.  The  envelopes (thick  curves) of  these two  flavors of
 droughts, represented  by their respective leading  harmonics ($\sim$
 120, 60, 40 days), reinforces this distinction.  The genesis of these
 ``climatological  sub-seasonal  oscillations''  \cite{wang_xu_97}  is
 plausibly  linked to  a preferred  temporal clustering  of ``breaks''
 \textcolor{black}{across the  years in the respective  categories (red
   and blue ellipses in} \textcolor{red}{Fig. \ref{fig:fig_s2}}).

 The stark  difference in timing and  duration of the breaks  in these
 two categories is captured in  Fig. \ref{fig:fig_2}b, which shows the
 respective composites  of the cumulative of  daily rainfall anomalies
 (see \textcolor{red}{Fig. \ref{fig:fig_s3}} for individual years).  A
 comparison of these cumulatives also shows that NEN + Dr has a larger
 overall deficit  than EN  + Dr  till early  July and  recovers around
 mid-July.   While the  timing may  be different,  the rate  of change
 (drop) in both categories is about 2-2.5 mm/d (0 to -100 mm in $\sim$
 45 days during June-July for EN + Dr  {\it vs} -25 mm to -75 mm in 20
 days during mid-to-late August for NEN + Dr).  The final state of the
 two curves in Fig.  \ref{fig:fig_2}b is in accord with the fact that,
 on average, the  magnitude of the seasonal deficit is  larger in EN +
 Dr  (17\%)  than  in  NEN  +  Dr (13\%),  as  can  be  inferred  from
 \textcolor{red}{Table \ref{tab:tab_s1}}.  While the  discussion so far has revolved
 around central India,  the distinction between the  two categories of
 droughts     holds      for     all     of     India      as     well
 (\textcolor{red}{Fig.  \ref{fig:fig_s4}}).  Thus,  an examination  of
 the  daily rainfall  evolution of  all droughts  in the  past century
 reveals  that  EN and  NEN  droughts  are seasonal  and  sub-seasonal
 phenomena, respectively.

Returning to the  original basis of our  classification, namely, SSTs,
while  EN +  Dr has  a well-defined  warm anomaly  extending into  the
central Pacific  (Fig.  \ref{fig:fig_1}a), the only  prominent feature
that stands  out in  the second category  (NEN + Dr)  is a  strong and
coherent  negative  SST anomaly  in  the  North Atlantic  ocean  (Fig.
\ref{fig:fig_1}b).   The  association  of  El  Ni\~{n}o  with  monsoon
droughts, characterized  by a  season-long rainfall deficit,  has been
extensively explored \cite{krishnakumar2006,turner2012,li2015}.  Here,
we focus  on the latter  category (NEN  + Dr), hitherto  less explored
\cite{varikoden2014},  and provide  a mechanistic  explanation of  how
conditions  over the  North Atlantic  lead  to an  abrupt late  season
decline in rainfall over India, resulting in a sub-seasonal drought.

\begin{figure}
  \centering
\includegraphics[width=0.75\textwidth]{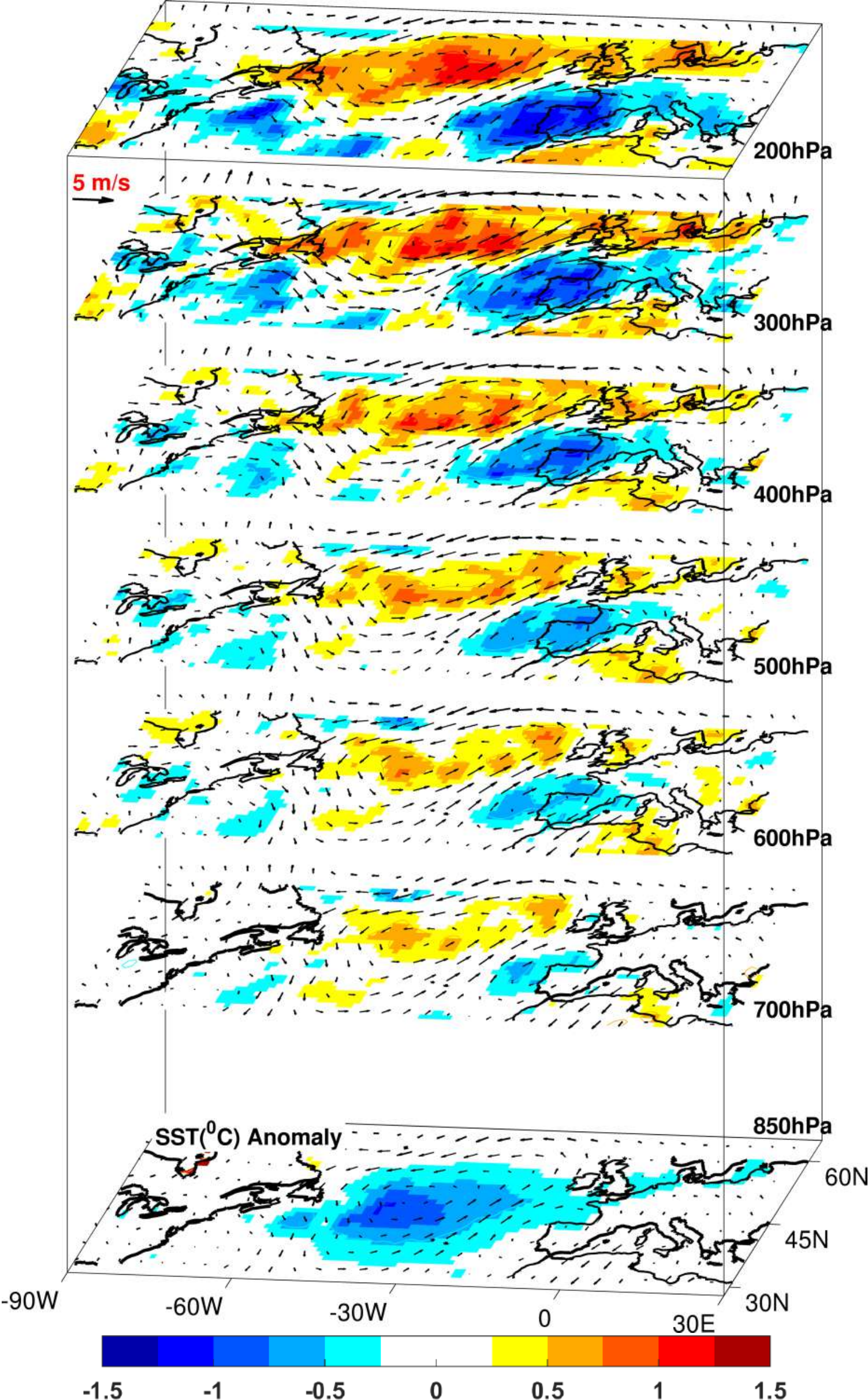}  
\caption{Composites  of  anomalies of  JJAS  mean  wind (vectors)  and
  vorticity  (shading) in  the free  troposphere,  based on  NEN +  Dr
  years; the shading in the  bottom-most panel represents SST anomaly.
  The colorbar applies to both  vorticity ($\times$ 2 $\times$ $10^5$)
  and SST.   The wind and  vorticity anomalies  are based on  ERA 20th
  Century Reanalysis data \cite{poli2016}, and SST anomalies are based
  on  the 1$^{\circ}$,  monthly  SST product  from  the Hadley  Centre
  \cite{hadsstref}.   See  Methods  section for  the  construction  of
  composites.}
\label{fig:fig_3}
\end{figure}

\begin{figure}
  \centering
\includegraphics[width=0.9\textwidth]{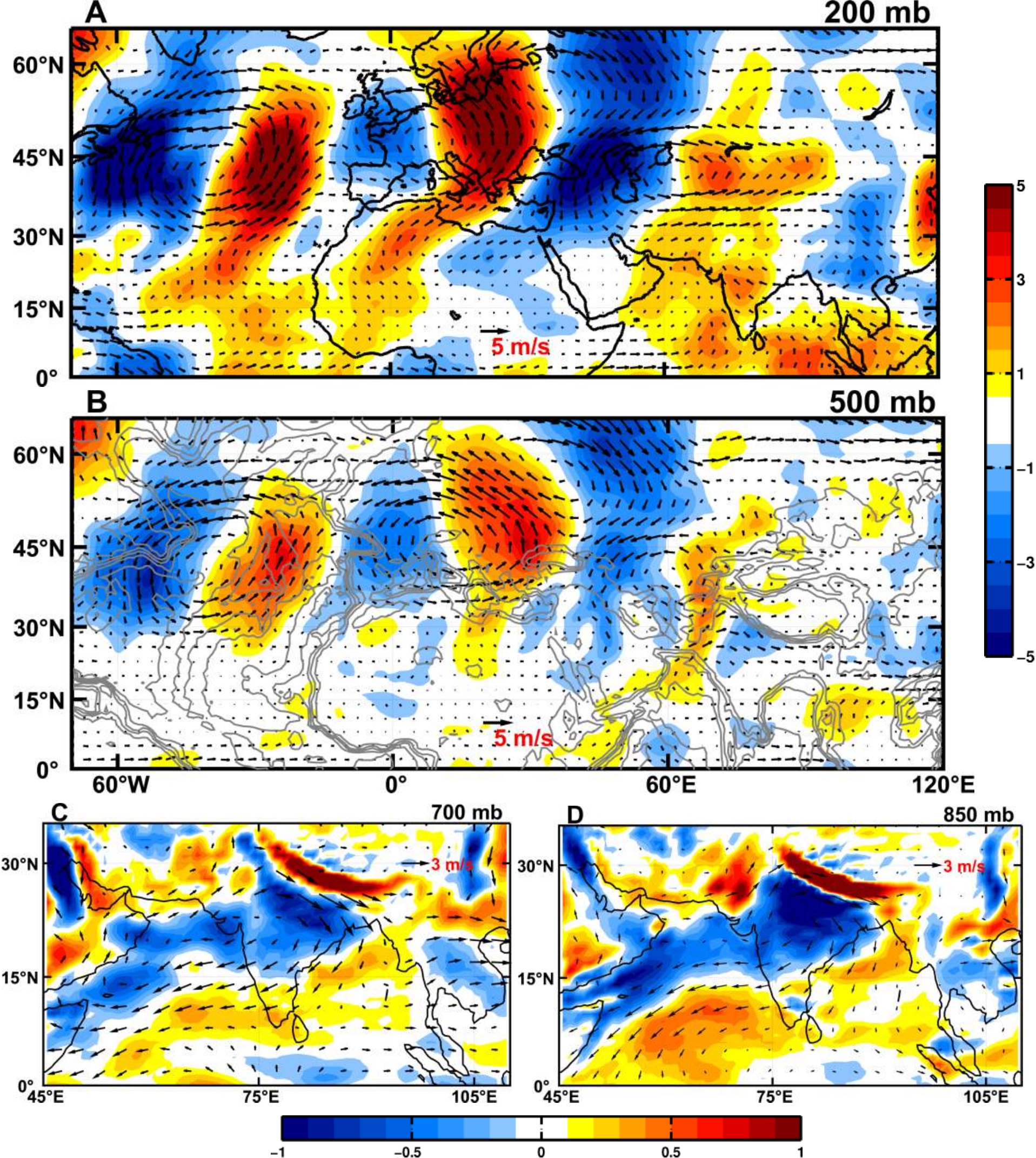}
  \caption{Composites  of   anomalies  of  (top  two   panels)  wind
      (vectors) and  meridional velocity (shading)  at (A) 200  mb and
      (B) 500 mb  during the first half of the  break (days 81-90; see
      blue  curve in  Fig.   \ref{fig:fig_2}b) in  the Indian  monsoon
      rainfall; and  (bottom two panels) wind  (vectors) and vorticity
      ($\times 10^5$; shading) at (C) 700  mb and (D) 850 mb, based on
      the respective break periods late in  the season for each of the
      NEN+Dr  years.  The  gray lines  in the  500 mb  panel represent
      topography.   Based   on  ERA   20th  Century   Reanalysis  data
      \cite{poli2016}.  See  Methods section  for the  construction of
      composites.}
      \label{fig:fig_4}
  \end{figure}

The seasonal  (JJAS) wind anomaly, averaged  for NEN + Dr  years, over
this North Atlantic oceanic marker, is shown in Fig.  \ref{fig:fig_3}.
The equivalent barotropic nature  of the regional atmospheric response
is evident and consists of a cyclonic circulation that extends all the
way from  850 mb, where  it is compact, to  200 mb, where  it expands,
intensifies and  becomes more zonal in  character.  The teleconnection
from the North Atlantic to the  Indian region is examined by analyzing
upper  tropospheric winds  during  the long  break  in Indian  monsoon
rainfall  (days  81-90;  see Fig.   \ref{fig:fig_2}b).   Specifically,
Figs.  \ref{fig:fig_4}a  and \ref{fig:fig_4}b show the  interaction of
the  upper  level flow  with  this  persistent and  deep  tropospheric
vorticity  forcing (Fig.   \ref{fig:fig_3}) during  this period.   The
resultant  200   mb  meridional   wind  anomalies  (shading   in  Fig.
\ref{fig:fig_4}a) indicate a response to the east of the Atlantic that
has  a   large  footprint   extending  all  the   way  to   East  Asia
\textcolor{black}{\cite{hoskins2000, branstator2017}}.  At 500 mb (Fig.
\ref{fig:fig_4}b),  as  expected,  the   anomalies  are  weaker;  more
interestingly,  the  response  curves  towards  the  equator,  with  a
prominent signature over  the Indian region, apparently  guided by the
Tibetan plateau  (the narrowing of the  wind anomalies to the  east of
60$^{\circ}$E in Fig. \ref{fig:fig_4}b).
The anticyclonic nature  of wind anomalies over North  India, signs of
which are evident  at 500 mb, becomes clearer at  lower levels (700 mb
and  850  mb;  Fig.   \ref{fig:fig_4}c  and  \ref{fig:fig_4}d).   This
circulation anomaly not  only weakens the zonal flow  over the Arabian
Sea  (i.e.,  disrupting the  Somali  Jet),  but also  strengthens  and
spreads to cover  all of India, as  we descend from 700 mb  to 850 mb.
Contrasting this  with the low-level cyclonic  circulation over India,
20 days prior to  the break (\textcolor{red}{Fig. \ref{fig:fig_s5}}),
highlights the abruptness of this phenomenon.

Our results  establish that  NEN droughts are  sub-seasonal and  are a
consequence of remote columnar  atmospheric forcing above a persistent
cold  North Atlantic  SST anomaly.   From a  broader perspective,  the
\textcolor{black}{association of the Atlantic basin with Indian summer
  monsoon has been explored, albeit  mostly on decadal and interannual
  timescales
  \cite{bng2006amo,lu2006,rajeevan2008,li2008,kucharski2009,krishna2016amo,sabeerali2019}.}
It  is worth  noting that  nearly all  NEN droughts  occur during  the
negative  phase of  the  Atlantic Multidecadal  Oscillation (AMO;  see
\textcolor{red}{Fig.   \ref{fig:fig_s6}}).  Even  though all  ``cold''
North  Atlantic  episodes  do  not necessarily  result  in  a  monsoon
drought, they cause a break during the monsoon following the mechanism
elucidated   above    (see   \textcolor{red}{Figs.   \ref{fig:fig_s7},
  \ref{fig:fig_s8} and ``Supplementary Text''}).

\textcolor{black}{While the teleconnection (wave train) from the North
  Atlantic  vorticity  forcing to  the  Indian  region appears  to  be
  robust, the  role of SST  in the  development of this  forcing needs
  further  investigation.    In  recent  years,  there   is  increased
  recognition  that  SST anomalies  can  trigger  a deep  tropospheric
  response       in      the       midlatitudes      \cite{minobe2008,
    smirnov2015,willis2016}.   To   better  understand  this   in  our
  context,  we examine  the  total columnar  vorticity  in years  with
  negative  and   positive  SST   anomalies  in  the   North  Atlantic
  \textcolor{red}{(Fig. \ref{fig:fig_s9}a)}.  It is quite evident from
  \textcolor{red}{Fig. \ref{fig:fig_s9}b}  that there is  a propensity
  for a  cyclonic (anti-cyclonic) environment over  the North Atlantic
  in  years when  the  SST anomaly  is  negative (positive).   Equally
  importantly,  even though  the  SST anomaly  stays negative  through
  JJAS, the columnar vorticity anomalies are episodic, with a tendency
  to  occur  in  June  and  August,  and persist  for  2  to  3  weeks
  \textcolor{red}{(Fig.  \ref{fig:fig_s9}b)}.  As  would be  expected,
  vorticity anomalies during NEN + Dr  years, which have on an average
  colder   anomaly,  show   a  similar   build  up   (blue  curve   in
  \textcolor{red}{Fig.   \ref{fig:fig_s9}b}),    but   are   amplified
  significantly (see \textcolor{red}{Supplementary Text and Table S3})
  in August  \cite{jjas_vort}.  Thus, while  SST anomalies may  not be
  solely responsible, they seem to be  essential in the formation of a
  transient deep tropospheric circulation in this region.}

Taken  together, all  Indian monsoon  droughts can  thus be  linked to
slowly-varying oceanic markers: the well-known  one in the tropics (El
Ni\~{n}o)  and  the   one  explored  in  this  work,   namely  in  the
midlatitudes (North Atlantic).  It is important to note that, in both
categories the respective  markers appear to be necessary  but are not
sufficient conditions  for the  occurrence of  a drought.   That these
drivers are slowly varying offers  an avenue for potential improvement
in predictability of  droughts. 
While the influence of midlatitudes on the monsoon circulation \cite{SchBor}
and   ISMR   variability    \cite{krishnan2009,yadav2009}   has   been
recognized, the sub-seasonal  mechanistic atmospheric bridge unraveled
in   this  study   represents   a  substantial   improvement  in   our
understanding  of connections  between the  North Atlantic  and Indian
monsoon  droughts.   More  pertinently,  it points  to  the  need  for
expanding the  existing tropics-centered paradigm of  monsoon droughts
into   a  framework   that   includes  extratropical   teleconnections
\cite{stan2017}.

\clearpage

\bibliographystyle{naturemag}

\clearpage


\clearpage

\clearpage

\clearpage

\clearpage

\clearpage


\setcounter{figure}{0}
\setcounter{table}{0}
\thispagestyle{empty}

\centerline{\Large {\bf Supplementary Material}}

\setcounter{page}{1}

\noindent{\large{\bf Data and Methods}}

\vspace*{0.2in}

\noindent
{\bf Data}
\begin{itemize}
\item  {\it Rainfall:}  1901-2015, daily, 1$^{\circ}$ gridded rainfall
    dataset from the India Meteorological Department (IMD). This
    product can be obtained from {\tt www.imd.gov.in}.

\item   {\it Sea Surface Temperature (SST):} 1901-2015, monthly, 1$^{\circ}$ gridded Had-SST from the Hadley Centre, UK Met
  Office \cite{hadsstref}. This dataset can be   obtained from\\
  {\tt https://www.metoffice.gov.uk/hadobs/hadsst3/data/download.html}.

\item {\it ERA  20C}: 1901-2010, daily, 1$^{\circ}$  gridded zonal and
  meridional winds and vorticity. This product \cite{poli2016} can be obtained from
  
  {\tt
    https://www.ecmwf.int/en/forecasts/datasets/reanalysis-datasets/ era-20c}.
  
\end{itemize}

\noindent{\bf Methods}

\begin{itemize}

\item  {\it Identifying  El Ni\~{n}o  Years}: El  Ni\~{n}o years  were
  identified by an inspection of  anomalies of detrended SST during JJAS.
  Post-1950 events were confirmed based  on estimates from the Climate
  Prediction Center, NOAA, and can be  accessed via ``El Ni\~{n}o / La
  Ni\~{n}a   $\longrightarrow$   Historical  Information''   at   {\it
    https://www.cpc.ncep.noaa.gov/}.

\item {\it Detrended  SST Anomalies:} A linear trend in  JJAS mean SST
  is  removed  at   each  location  for  the   period  1901-2015,  and
  fluctuations around the trend composited for the years of interest.
  
\item {\it  Rainfall, Wind and Vorticity  Anomalies:} If $F(x,y,d,yr)$
  represents rain, wind or vorticity at  a location $(x,y)$ on day $d$
  during JJAS, in year $yr$, then the daily anomalies at each location
  are estimated as:
  $F_{\rm anom}(x, y, d, yr) = F(x,y,d,yr) - F_{\rm clim}(x, y, d)$
  where
  $F_{\rm       clim}(x,       y,       d)       =       \frac{1}{N}
  \left[\sum\limits_{yr=1}^{yr=N} F(x,y,d,yr)\right]$.
  The anomalies of  {\it area-averaged rainfall} (e.g.,  Fig.  2a) are
  estimated  by  calculating  daily area-averaged  rainfall  and  then
  subtracting  the  corresponding  daily climatology.   {\it  Seasonal
    anomalies}  of $F$  (e.g., Fig.   3 or  Fig. 4)  can be  estimated
  either  by subtracting  JJAS means  from the  corresponding seasonal
  climatology, or by averaging the daily anomalies $F_{\rm anom}(x, y,
  d, yr)$ over JJAS.
  
\item {\it  Composites}: The  ERA20C data is  available up to  2010 and
  hence 2014  was not included  in the wind and  vorticity composites.
  Furthermore, the composites presented have been constructed based on
  the anomalies of 7 of the 9  NEN + Dr years for which the reanalysis
  data  is  available.  This  is  because  the  lower level  (850  mb)
  circulation features over the Indian  region are not consistent with
  the observed  IMD rainfall during the  break periods for two  of the
  early era  years (1901 and 1904).   This could be either  because of
  the  quality  of early  era  IMD  observations  or the  fidelity  of
  reanalysis   data  (the   years  being   close  to   spin-up  period
  \cite{poli2016}).
\end{itemize}

\clearpage

\noindent
{\Large{\bf Supplementary Text}}

\bigskip

\noindent
{\bf Intra-category Heterogeneity.} The rainfall evolution for each of
the droughts  suggests considerable intra-category  heterogeneity. The
spread  in normal  years (grey  dashed lines  in \textcolor{red}{Figs.
  \ref{fig:fig_s3}b  and \ref{fig:fig_s3}d})  is substantially  lower
than    in   either    of   the    drought   categories.     Moreover,
\textcolor{red}{Fig.  \ref{fig:fig_s3}d}  indicates that NEN +  Dr can
be near-normal until  as late as end of July/early  August, before the
abrupt transition to an eventual drought.\\

\noindent
{\bf Anomalously Cold  North Atlantic and No  Indian Monsoon Drought.}
It is interesting to note that  the evolution of the cumulative curves
for  the   two  categories,  namely   \{NEN  +  Droughts\}   (blue  in
\textcolor{red}{Fig.   \ref{fig:fig_s7}b})  and \{North  Atlantic  SST
$<0$    but    no     drought\}    (cyan    in    \textcolor{red}{Fig.
  \ref{fig:fig_s7}b})  are  nearly  the same  until  mid-August.   The
circulation features resulting in an  early season (June) break in the
latter category  are shown in  \textcolor{red}{Fig. \ref{fig:fig_s8}}.
The June break  mechanism is very similar to that  seen in mid-to-late
August for NEN  + Dr (Fig.~4).  In other words,  a cool North Atlantic
sets up a  vorticity forcing and its interaction  with the upper-level
flow  (\textcolor{red}{Fig.   \ref{fig:fig_s8}a})  results in  a  wave
train that  eventually leads to  an anticyclonic circulation  over the
Indian land mass (\textcolor{red}{Fig.  \ref{fig:fig_s8}c}).  Thus, it
appears that {\it two breaks} are  needed during a cold episode of the
North Atlantic for the Indian monsoon  to go into a drought-like state
(see also \textcolor{red}{Fig. \ref{fig:fig_s9}}).\\

\noindent
\textcolor{black}{The  {\bf  statistical  significance} of  JJAS  daily
  total  columnar  vorticity  anomalies  during  NEN  +  Dr  years  is
  established via a suite  of three tests (\textcolor{red}{Table S3}).
  The  first  test  shows   that  these  anomalies  are  significantly
  different from the long-term mean (zero by definition) in a seasonal
  sense with a $p$-value $\approx 10^{-7}$. The second and third tests
  are  more stringent  and involve  a comparison  with years  when the
  North  Atlantic   is  ``cold''  (SST  anomaly   $<  -0.25^{\circ}$C;
  \textcolor{red}{Fig.  \ref{fig:fig_s9}a}) and  there is a preference
  for      a      cyclonic      environment      (\textcolor{red}{Fig.
    \ref{fig:fig_s9}b}).  Here  too, the columnar  vorticity anomalies
  are  significantly  different both  in  a  seasonal  as well  as  an
  episodic sense, with $p$-values of 4.6\% and 0.3\%, respectively.}

\clearpage

\renewcommand\thefigure{S\arabic{figure}}
\renewcommand\thetable{S\arabic{table}}

\centerline{\large{\bf SUPPLEMENTARY FIGURES}}

\begin{figure}[h]
\begin{center}
  \includegraphics[width=150mm]{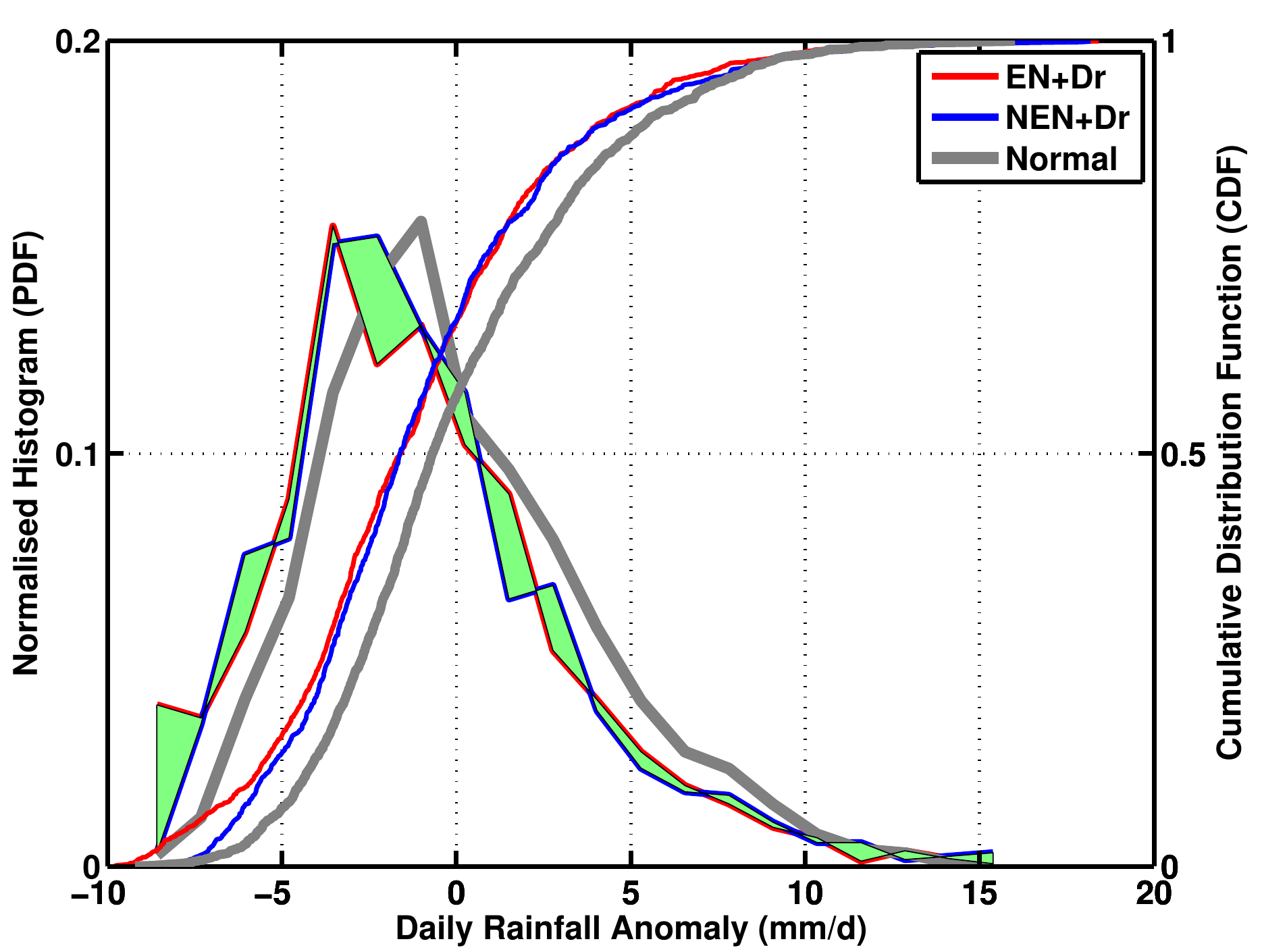}
  \caption{Normalized histograms (shown as line curves) and cumulative
    distribution functions  (CDF) of anomalies of  daily area-averaged
    rainfall over central India  (16.5N-26.5N, 74.5E-86.5E; land only;
    also, see also Methods section) for EN + Dr (red), NEN + Dr (blue)
    and  normal  (gray).   The   light-green  shading  highlights  the
    difference  between the  two kinds  of droughts  as well  as their
    difference between  them and a normal  year.  The CDFs of  EN + Dr
    (red) and NEN  + Dr (blue) are  statistically indistinguishable at
    5\%  significance   level  ($p$-value  of  0.09),   but  are  both
    significantly  different from  a  normal year  (gray ``S''  curve;
    $p$-value  $< 10^{-5}$).   The empirical  CDFs have  been compared
    using a two-sample Kolmogorov-Smirnov  test with a null hypothesis
    that they are indistinguishable \cite{ross2009}.}
\label{fig:fig_s1}
\end{center}
\end{figure}

\clearpage

\begin{figure}[h]
\begin{center}
{\includegraphics[width=100mm]{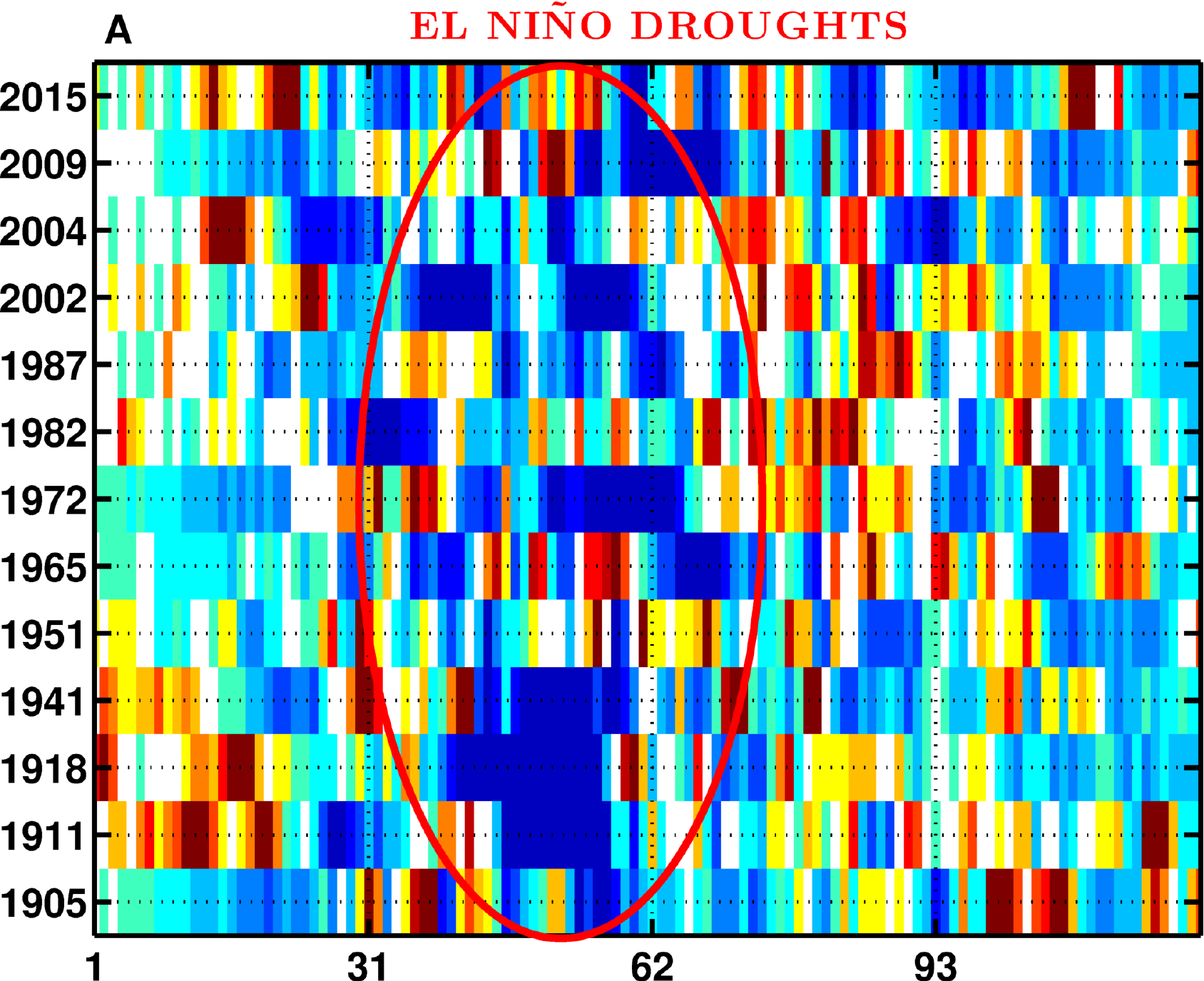}}

\vspace*{0.15in}

{\includegraphics[width=100mm]{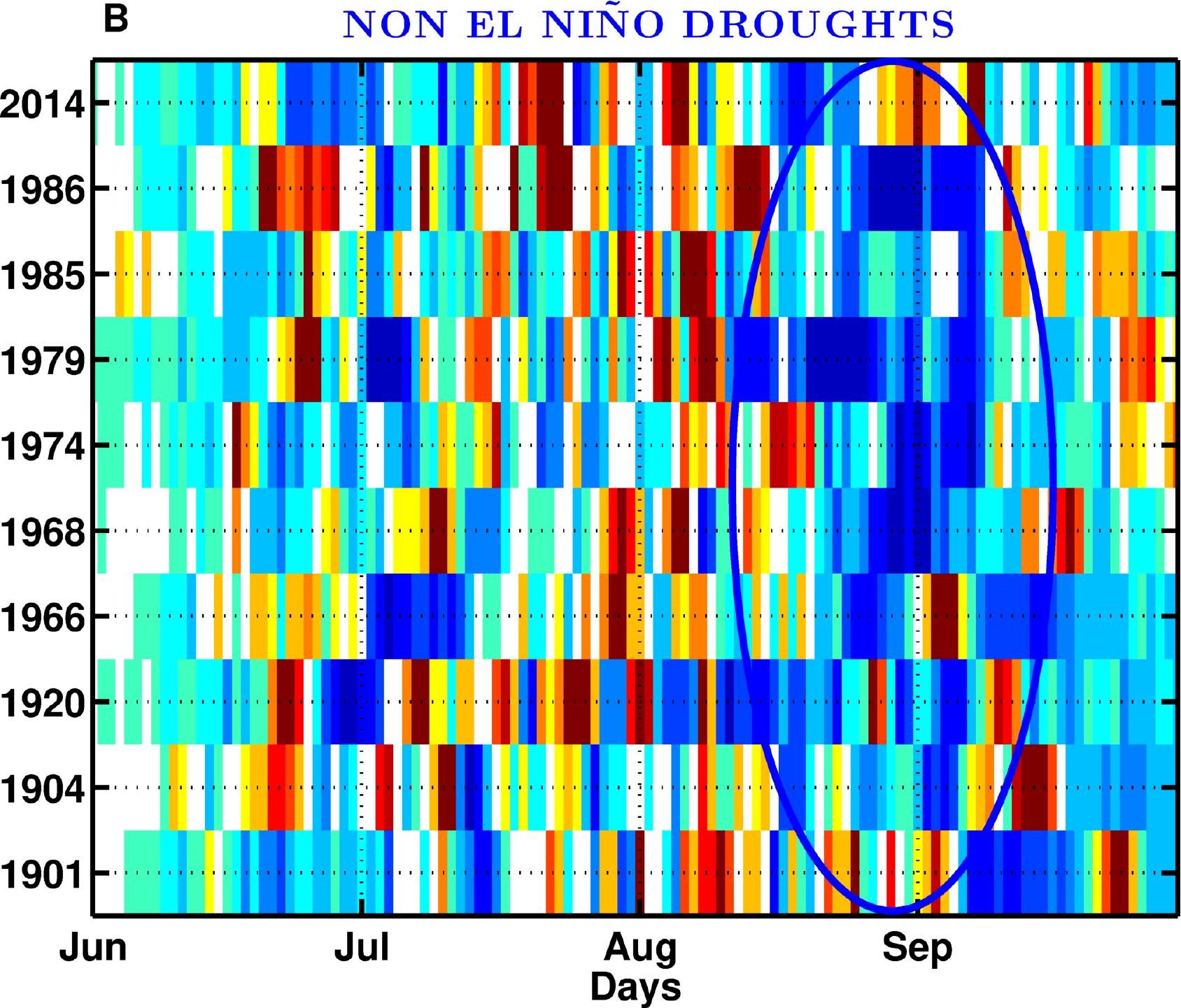}}

\medskip

\hspace*{4mm}  \includegraphics[width=80mm]{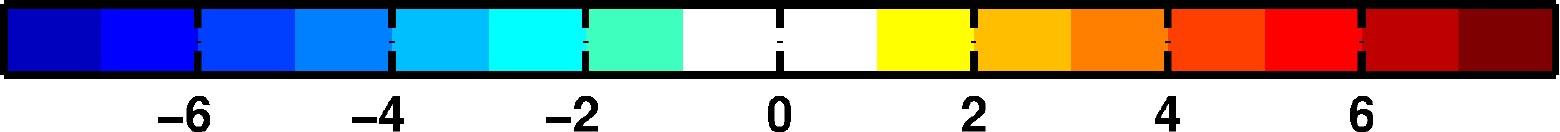} 
  \caption{Time series  of anomalies  of daily  area-averaged rainfall
    (mm/d) over  central India  (16.5N-26.5N, 74.5E-86.5E;  land only;
    see also Methods section), during JJAS, shown with a color-coding,
    for each of the drought years during (A) an El Ni{\~n}o year (EN +
    Dr) and (B) a  non El Ni\~{n}o year (NEN + Dr).   The red and blue
    ellipses  are used  to highlight  the tendency  of long  breaks to
    cluster in  different times  of the  season, in  the two  kinds of
    droughts. Colorbar is in mm/d. The average of these time series is
    shown   as   thin   red   and  blue   curves,   respectively,   in
    Fig.~2a.}
\label{fig:fig_s2}
\end{center}
\end{figure}

\clearpage

\begin{figure}[h]
\begin{center}
{\includegraphics[height=62mm]{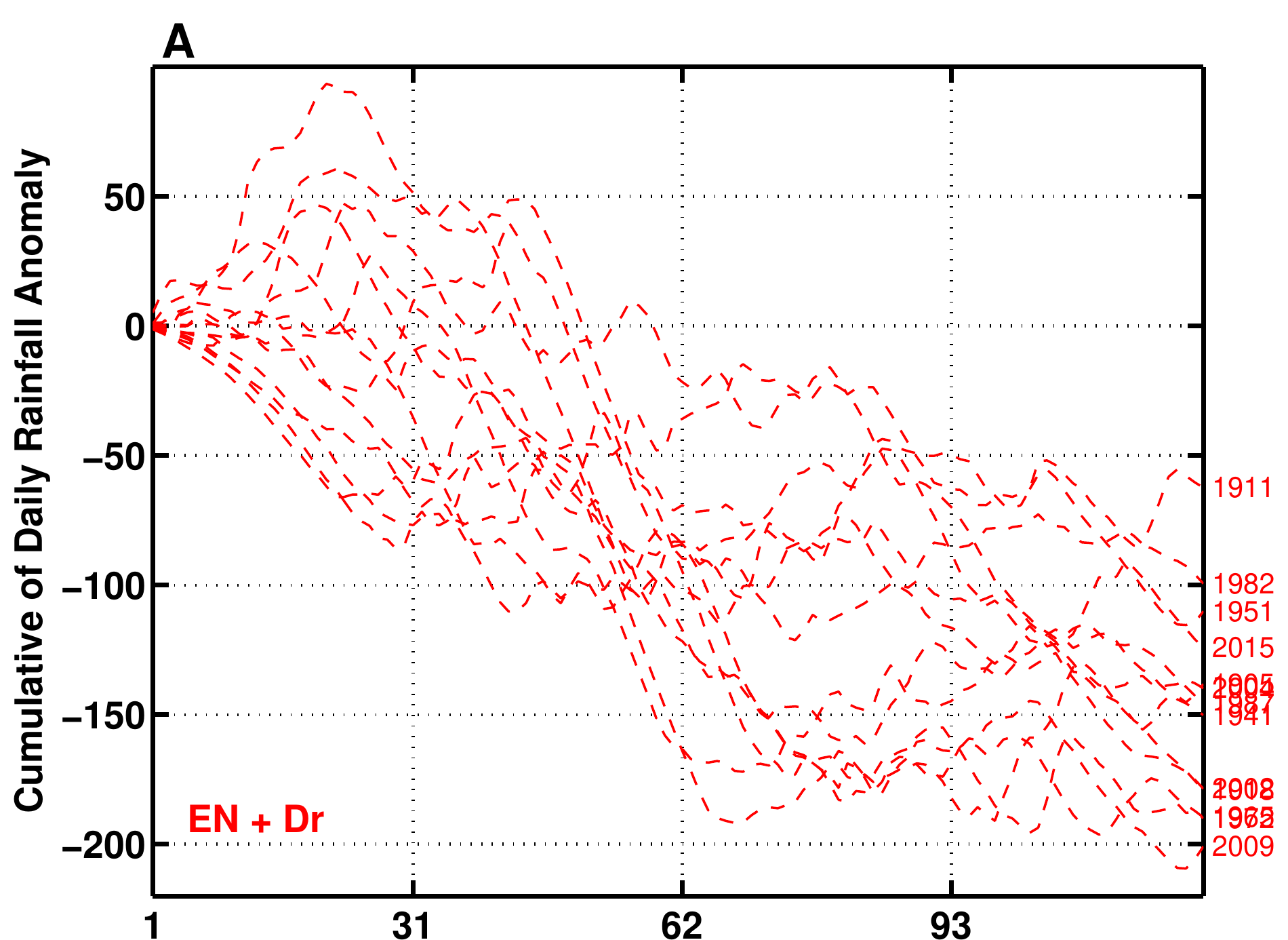}}
{\includegraphics[height=62mm]{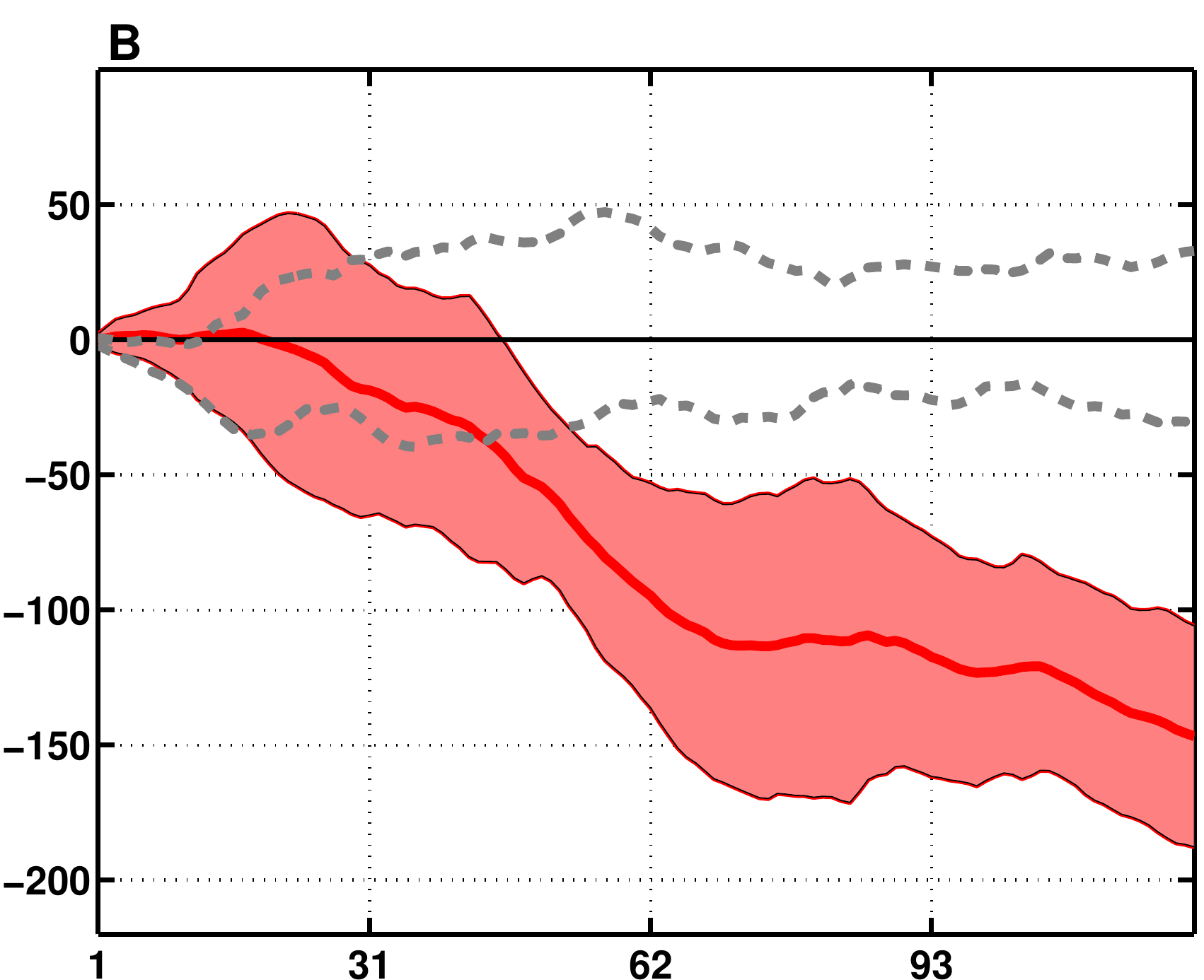}}

\medskip

{\includegraphics[height=62.5mm]{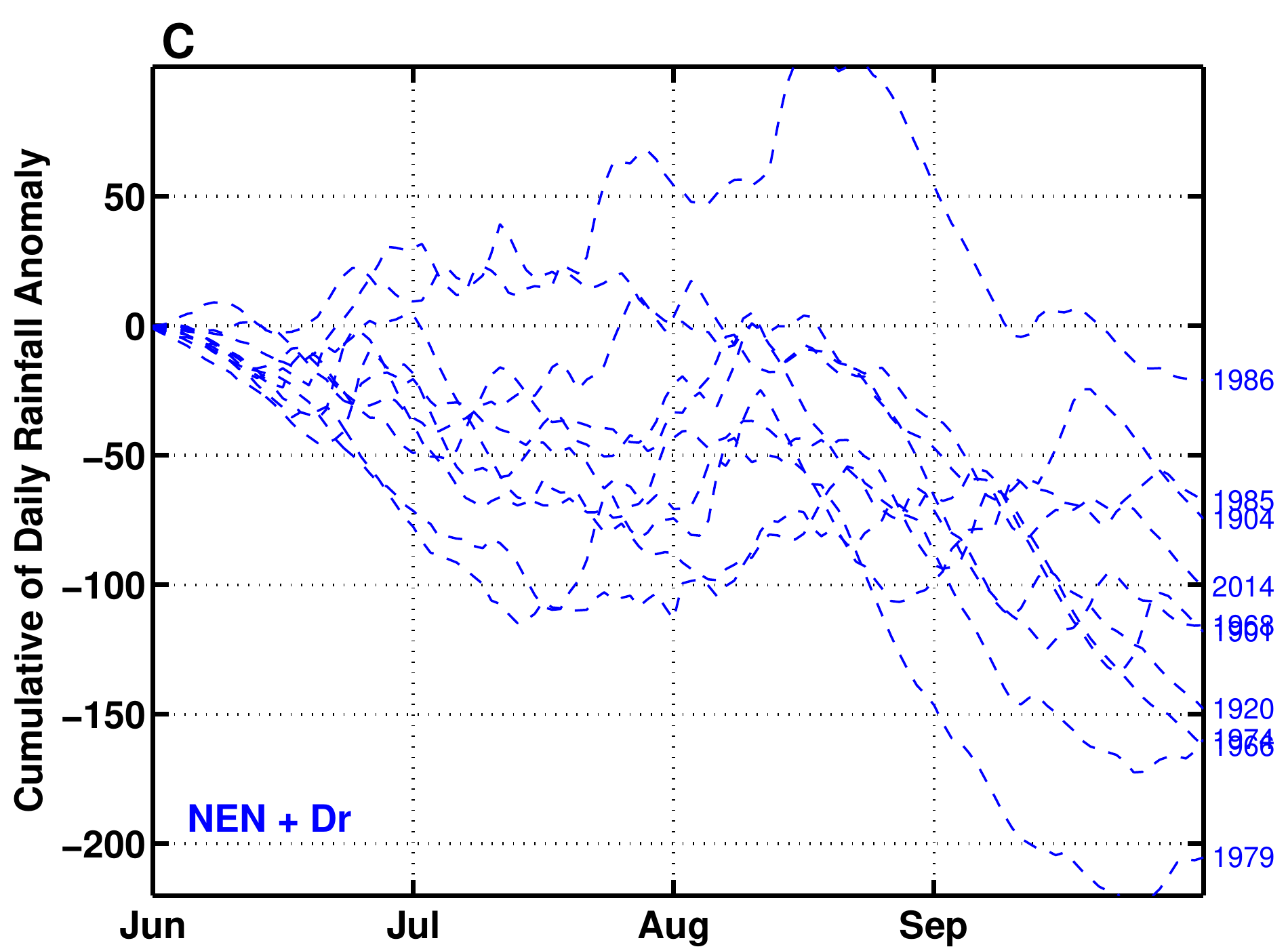}}
{\includegraphics[height=62.5mm]{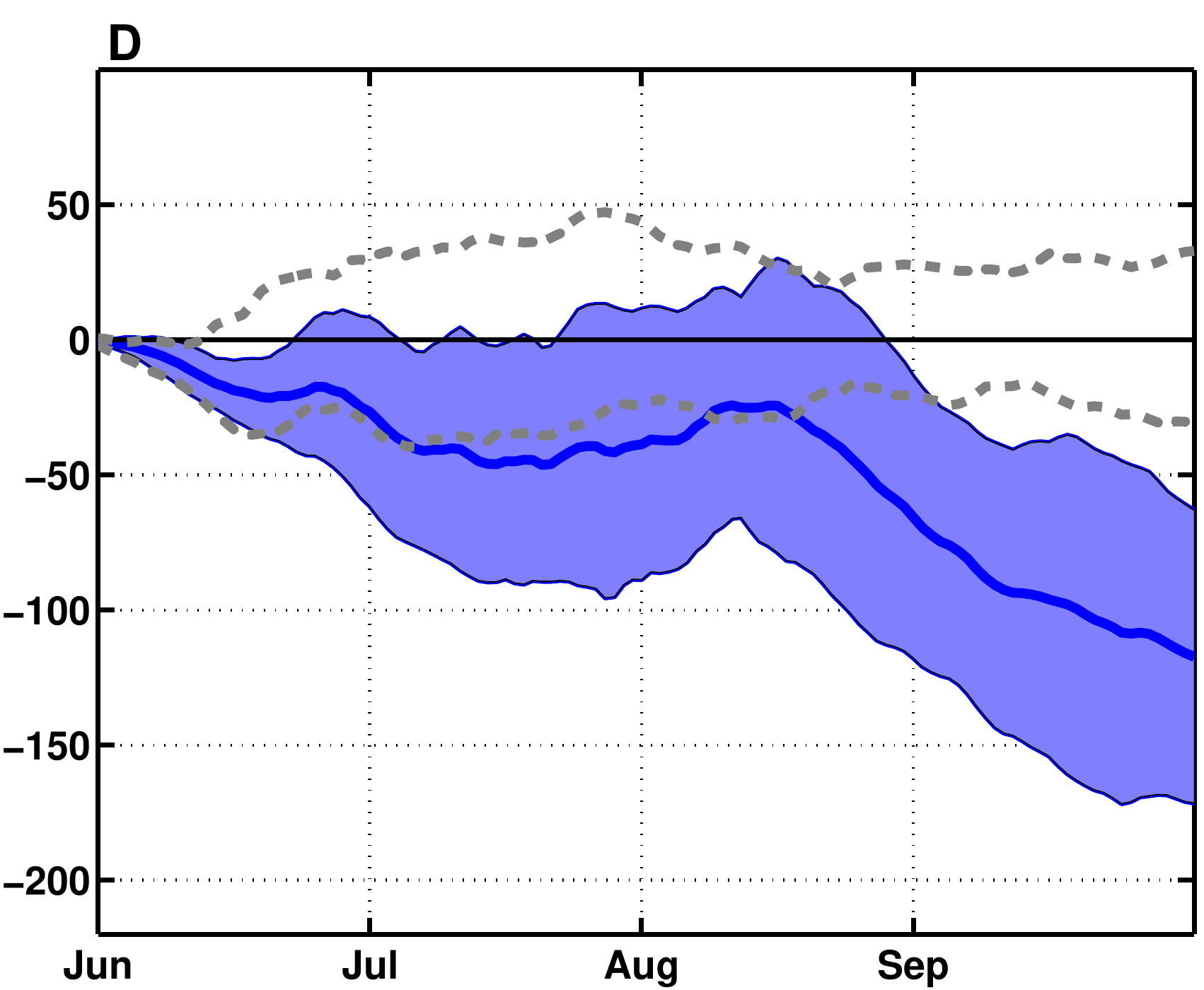}}
\caption{Cumulative  curves   of  daily  anomalies   of  area-averaged
  rainfall over  central India  (16.5N-26.5N, 74.5E-86.5E;  land only;
  see also Methods section) for each of the drought years belonging to
  the  two categories,  along with  the respective  average cumulative
  curves: (A,  B) EN  + Dr  and (C,  D) NEN  + Dr.   The corresponding
  spread  ($\pm$1 standard  deviation) in  each of  the categories  is
  shown as a shading in panels (B) and (D).  The standard deviation is
  estimated based  on 13 EN +  Dr and 10  NEN + Dr years.   The dashed
  gray lines  in panels (B) and  (D) correspond to the  spread ($\pm$1
  standard  deviation) of  18 normal  years.  The  solid red  and blue
  curves   in  (B,   D)   are  the   same  as   the   ones  shown   in
  Fig.~2b.}
\label{fig:fig_s3}
\end{center}
\end{figure}

\clearpage

\begin{figure}[h]
\begin{center}
{\includegraphics[width=120mm]{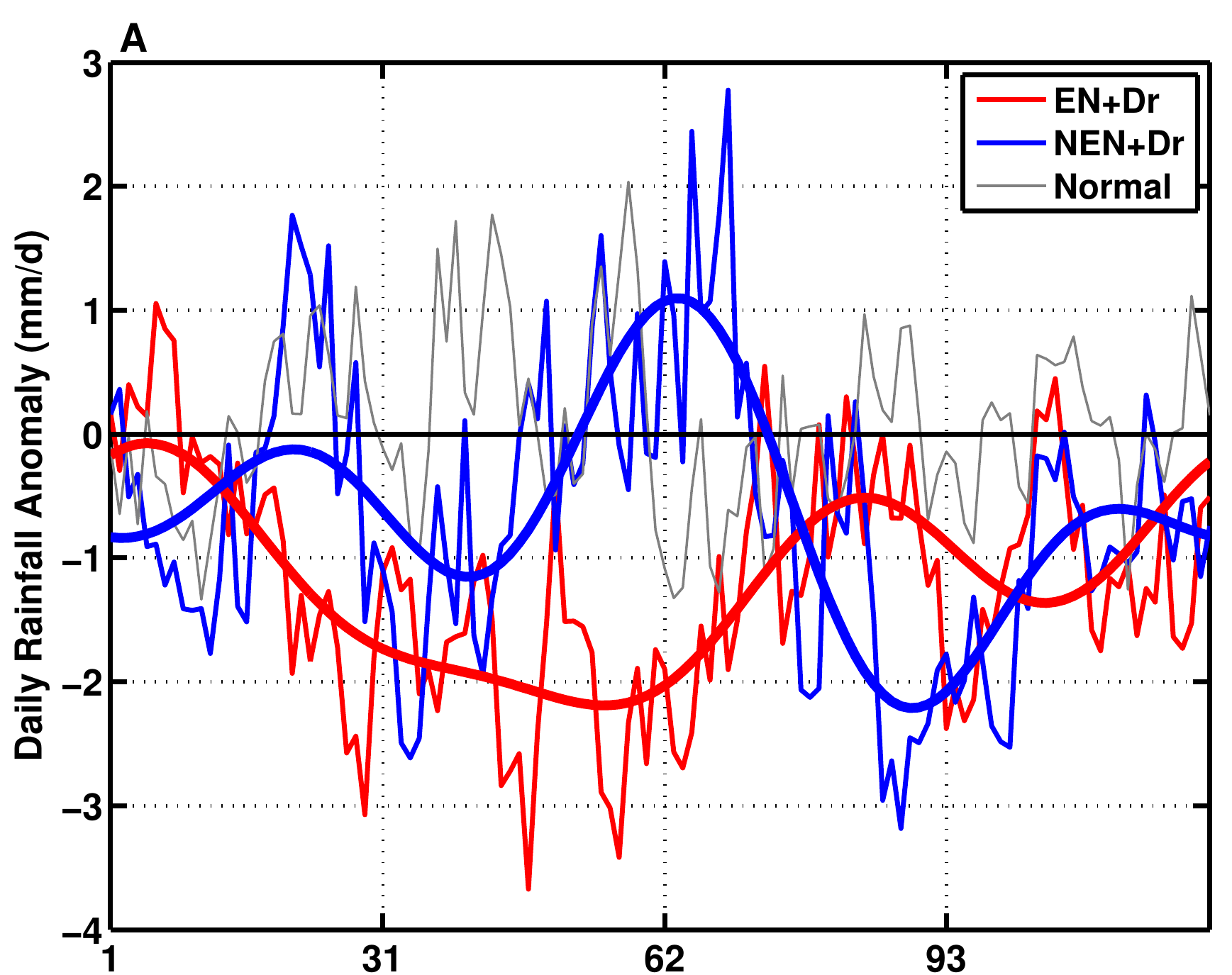}}
\hspace*{-6mm} {\includegraphics[width=123mm]{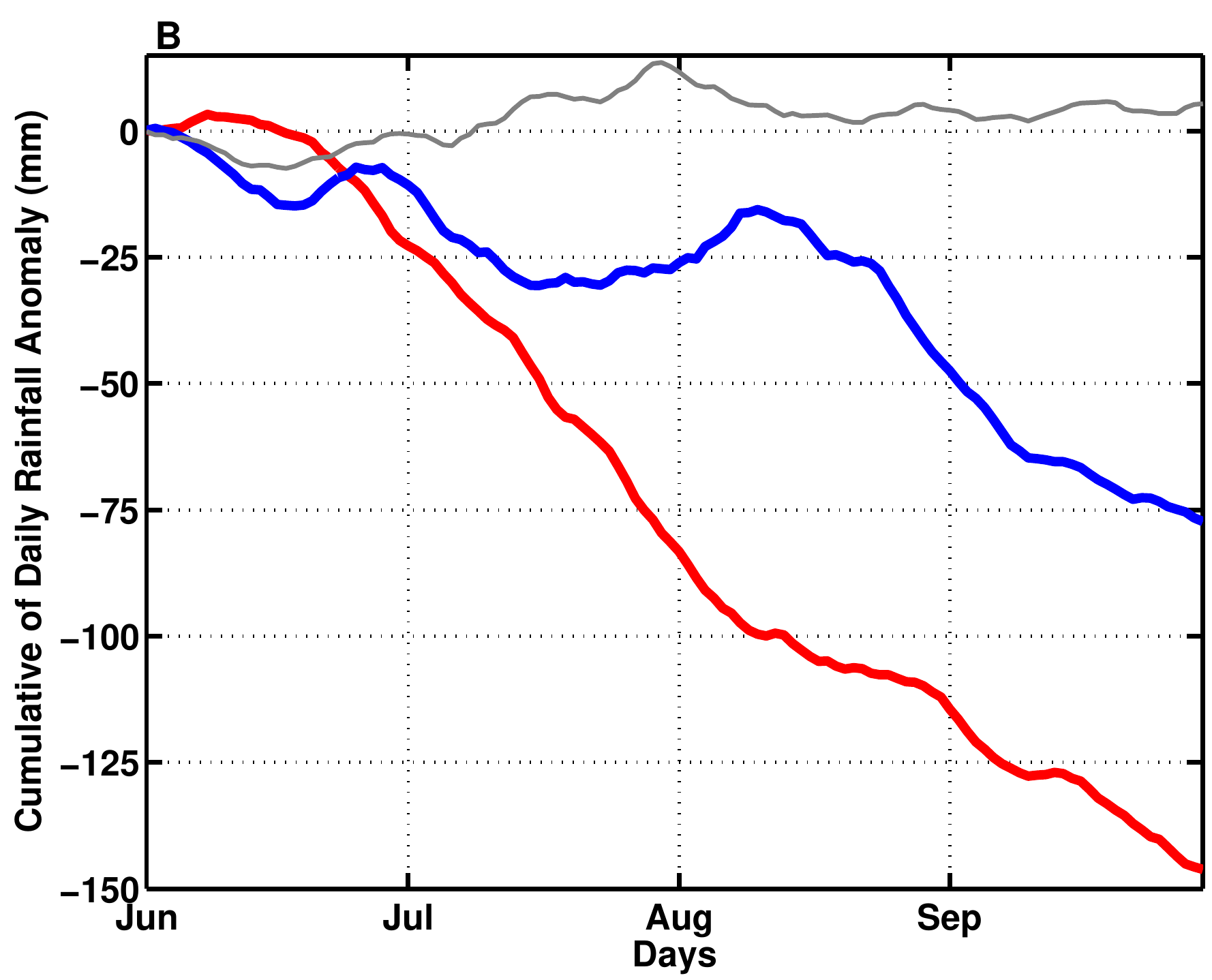}}
\caption{Same  as  in  Fig.~2,  but  for  area-averaged
  all-India rainfall.}
\label{fig:fig_s4}
\end{center}
\end{figure}

\clearpage

\begin{figure}[hbtp]
  \begin{center}
    \includegraphics[width=5.35in]{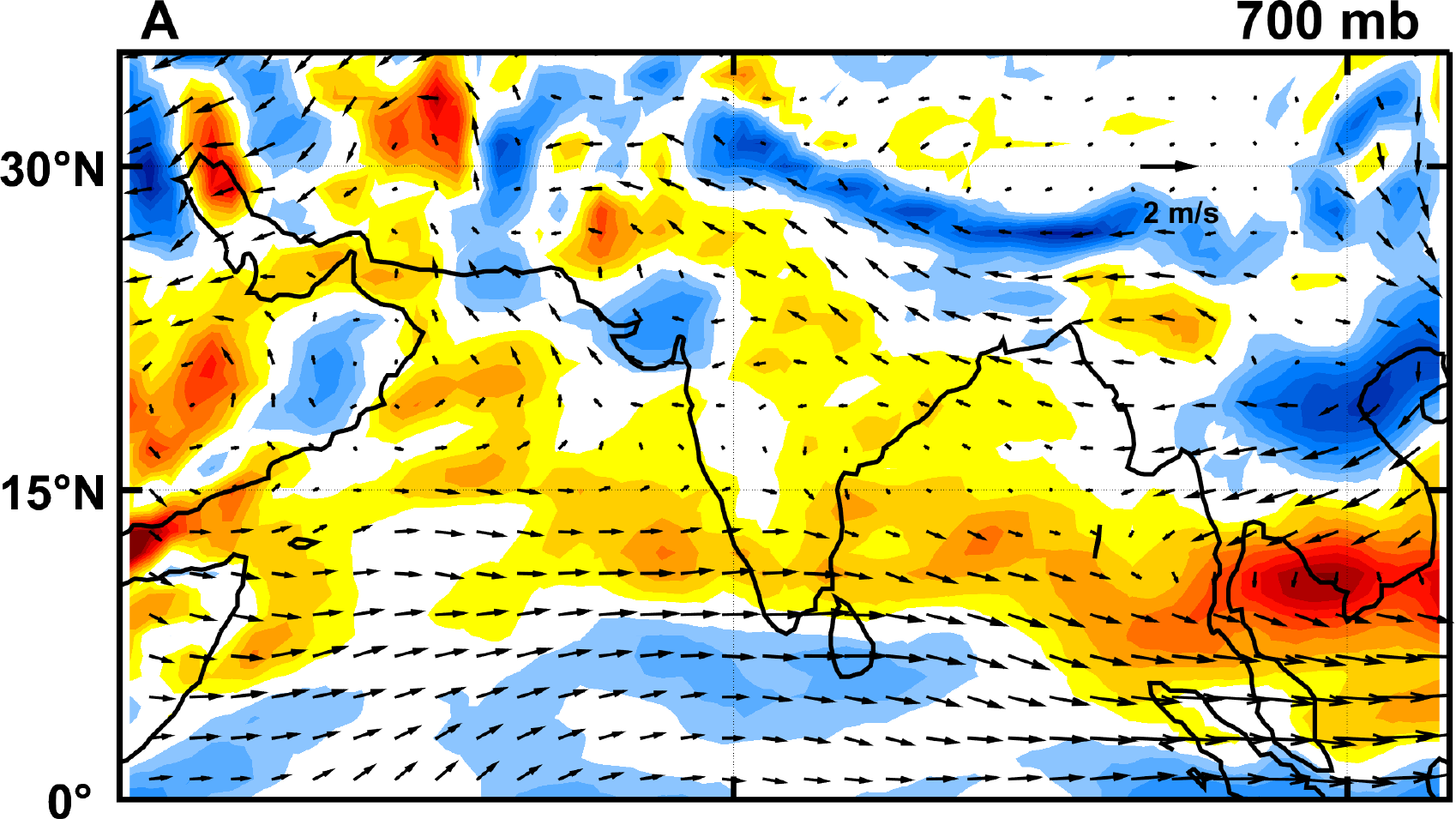}

    \includegraphics[width=5.35in]{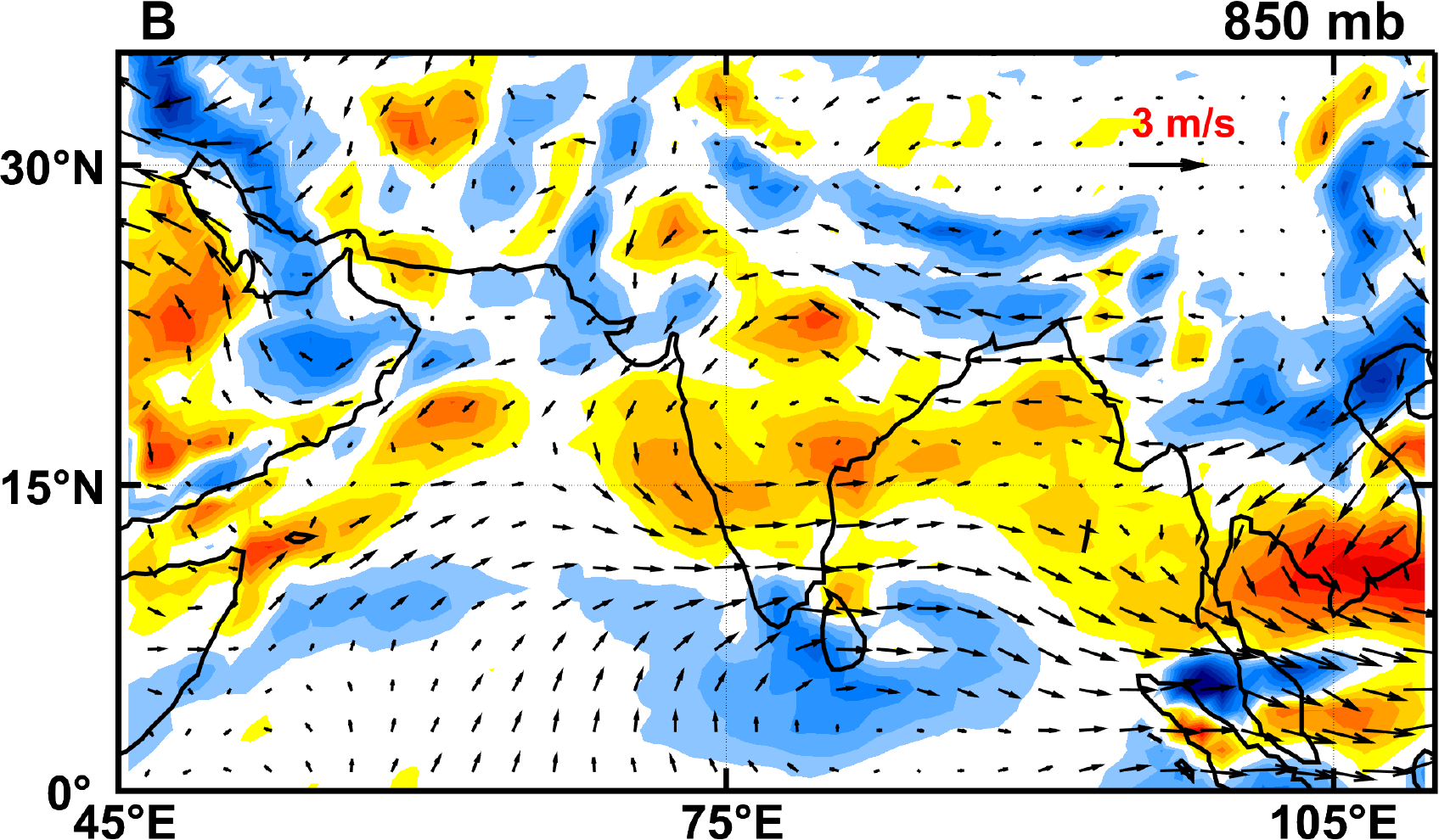}
    
 \includegraphics[width=90mm]{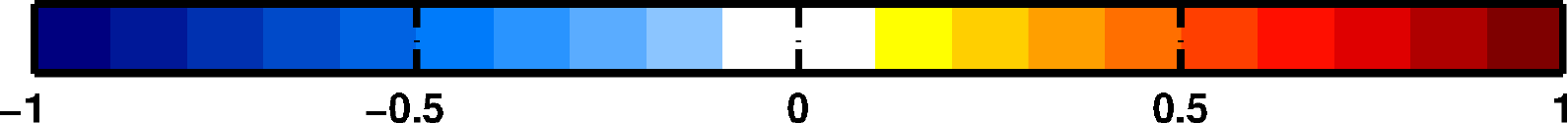}
    \caption{Composites of  anomalies of wind (vectors)  and vorticity
      ($\times 10^5$;  shading) at (A) 700  mb and (B) 850  mb, during
      the 20-day period  prior to the long break (see  blue ellipse in
      \textcolor{red}{Fig.    \ref{fig:fig_s2}B})  in   NEN+Dr  years.
      Legend for  arrows is  shown in  panel (B).   Based on  ERA 20th
      Century  Reanalysis data  \cite{poli2016}.  See  Methods section
      for the construction of composites.}
      \label{fig:fig_s5}
  \end{center}
  \end{figure}

\clearpage

\begin{figure}
\begin{center}
\includegraphics[width=150mm]{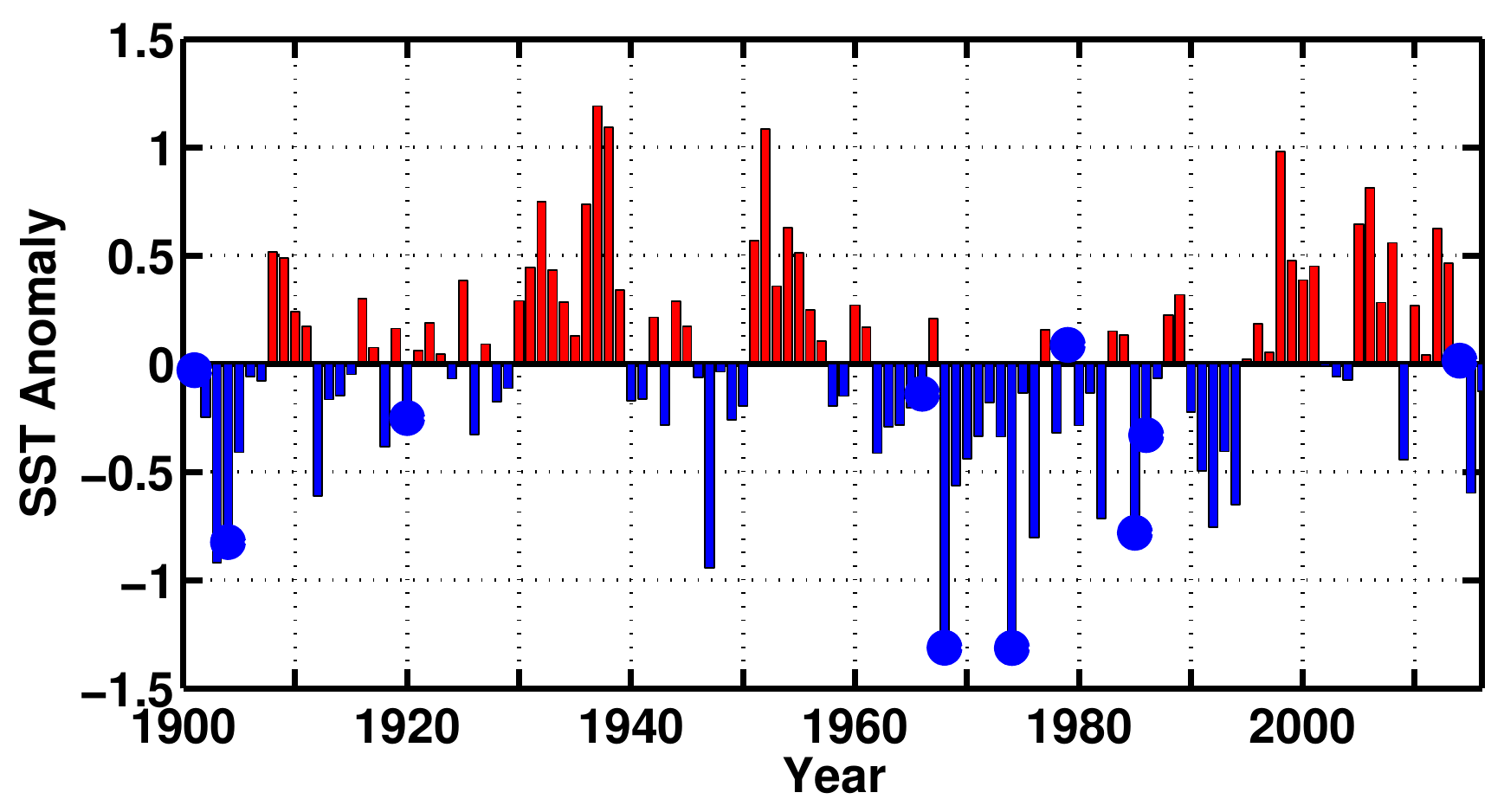}
\caption{Long-term  (1900-2015)  time  series of  the  North  Atlantic
  (35N-55N; 310E-340E; based on  Fig.~1b) SST anomalies
  ($^{\circ}$C) with  the 10 NEN  drought years marked as  blue filled
  circles.  Based  on the  1$^{\circ}$, monthly  SST product  from the
  Hadley Centre \cite{hadsstref}.}
\label{fig:fig_s6}
\end{center}
\end{figure}

\clearpage

\begin{figure}[h]
  \centering
\includegraphics[width=5in]{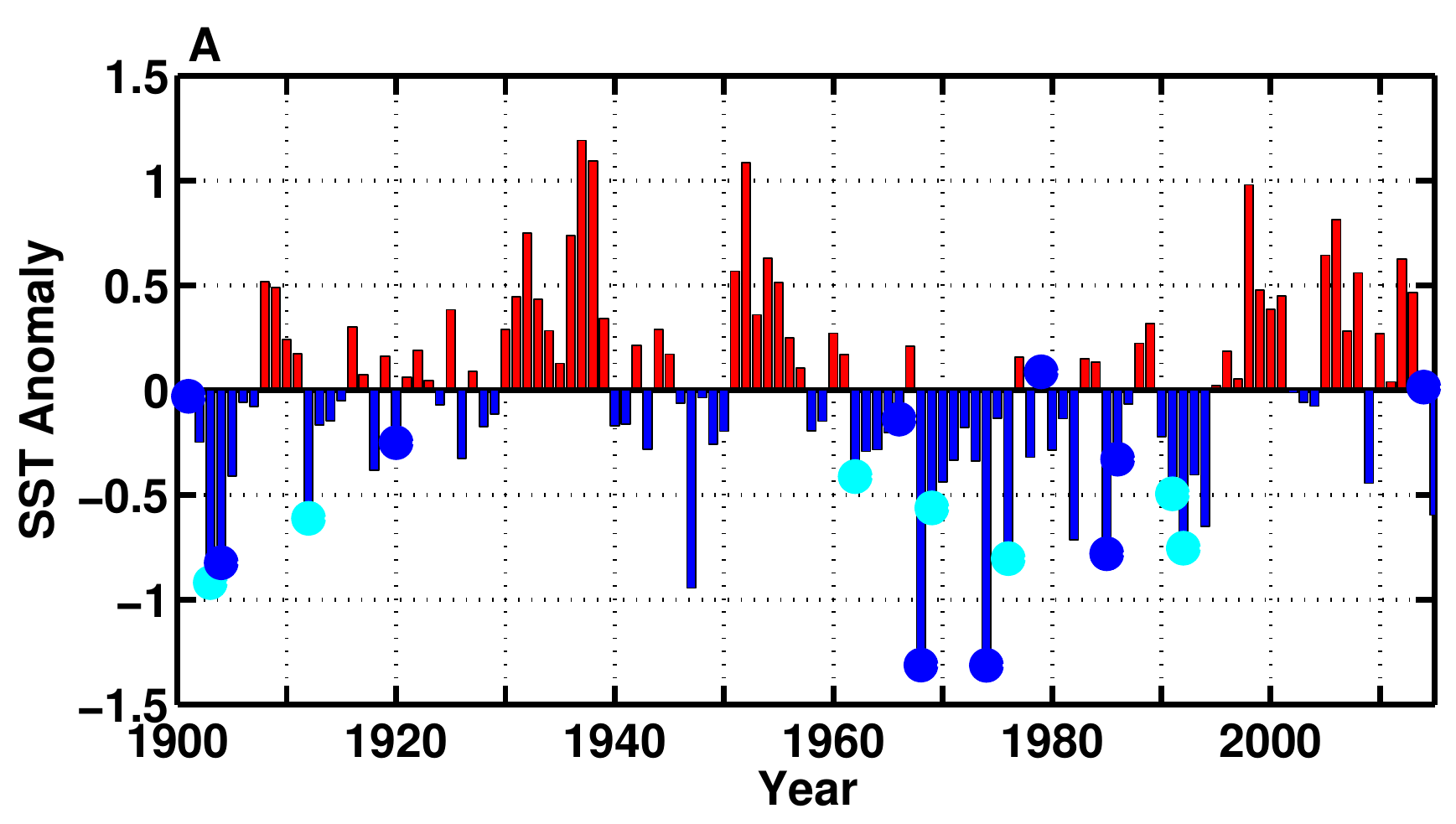}
  
\hspace*{-0.2in}\includegraphics[width=5.1in]{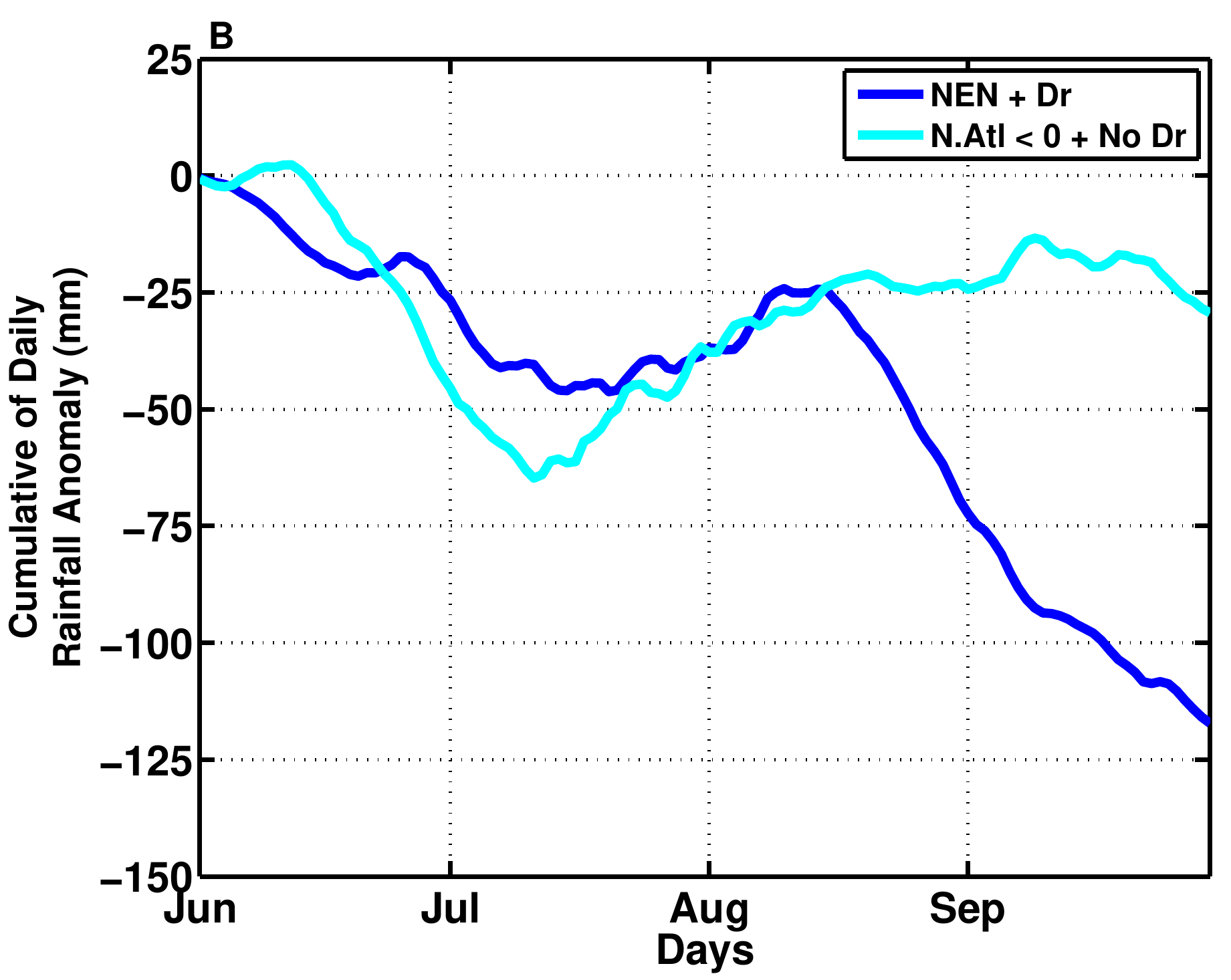}
\caption{(A)  Same as  in \textcolor{red}{Fig.   \ref{fig:fig_s6}} but
  also showing 7 no-drought years when the North Atlantic was negative
  as  cyan  filled  circles   (see  \textcolor{red}{Table  S2}).   (B)
  Cumulative of anomalies of daily area-averaged rainfall over central
  India (as in Fig.~2b)  for NEN + Dr years (blue  circles in (A)) and
  North Atlantic SST $<0$ $+$ no drought years (cyan circles in (A)).}
\label{fig:fig_s7}
\end{figure}

\clearpage

\begin{figure}[h]
  \centering
\includegraphics[width=6.25in]{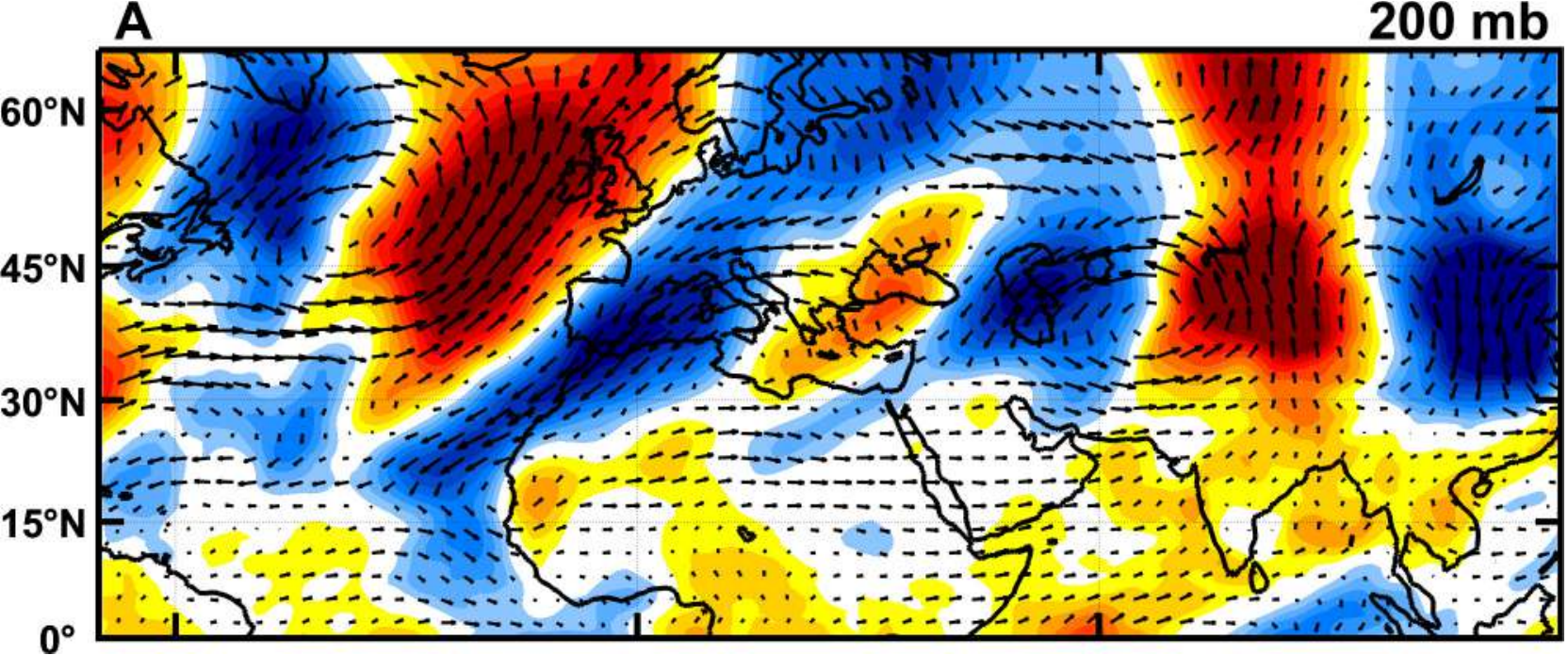}
\includegraphics[width=6.25in]{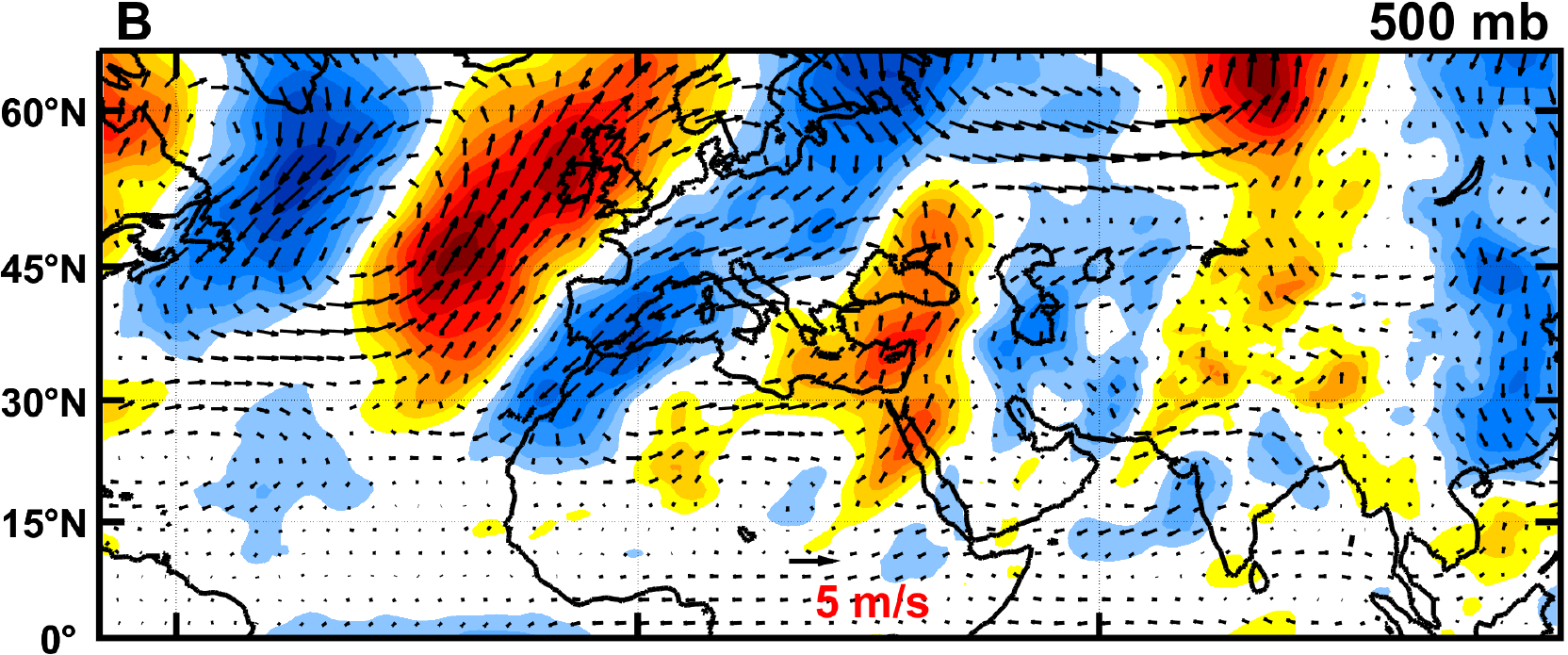}
\hspace*{0.1in}\includegraphics[width=6.5in]{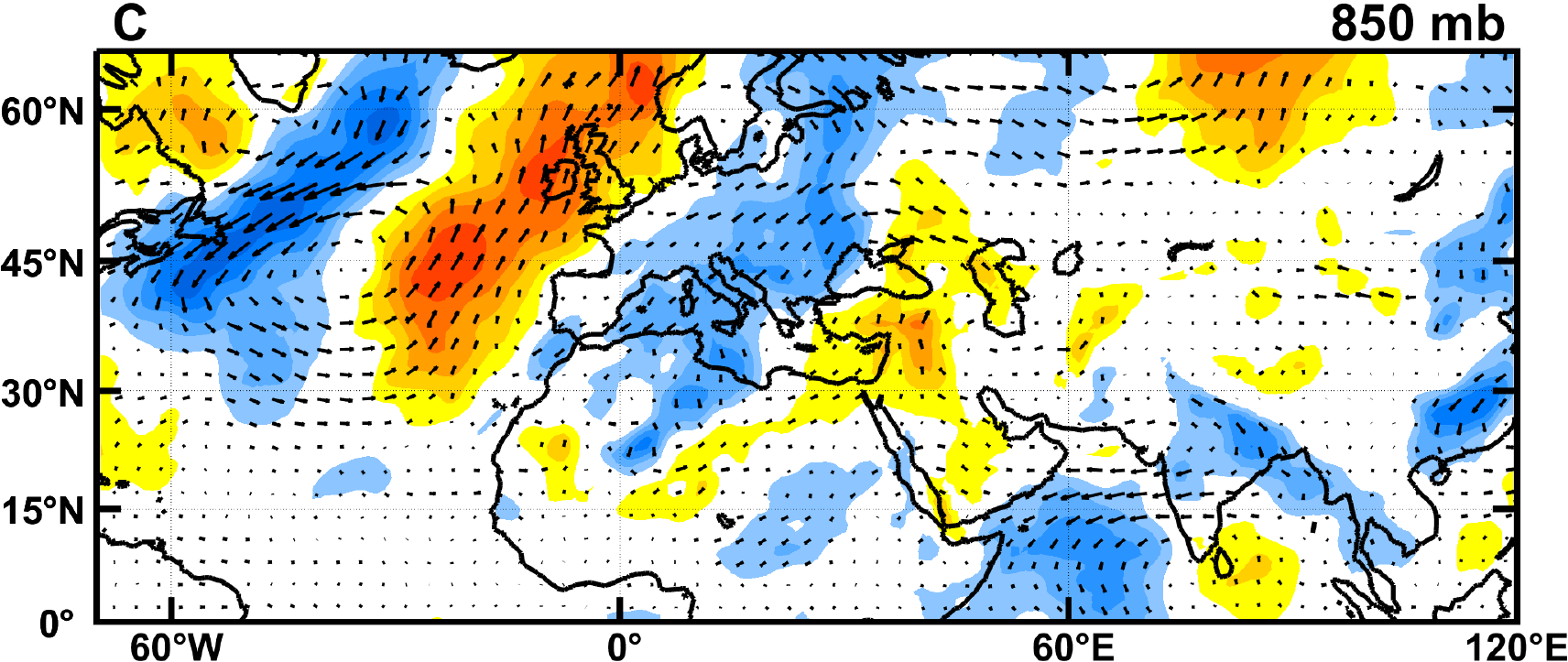}

\includegraphics[width=4.5in]{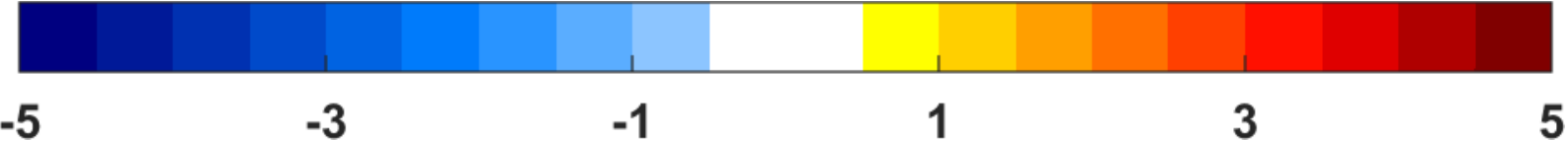}
\caption{Composite  of  anomalies  of wind  (vectors)  and  meridional
  velocity (shading) at (A) 200 mb; (B) 500 mb; and (C) 850 mb, during
  days   20-30   (i.e.,  Jun   20-30)   for   the  years   listed   in
  \textcolor{red}{Table  S2}  and  shown  as cyan  filled  circles  in
  \textcolor{red}{Fig.  \ref{fig:fig_s7}a}.   These days are  based on
  the early-season dip in the cyan curve shown in \textcolor{red}{Fig.
    \ref{fig:fig_s7}b}.  Legend for arrows is shown in panel (B).}
    \label{fig:fig_s8}
  \end{figure}

\clearpage

\begin{figure}[btph]
  \centering
\includegraphics[width=5in]{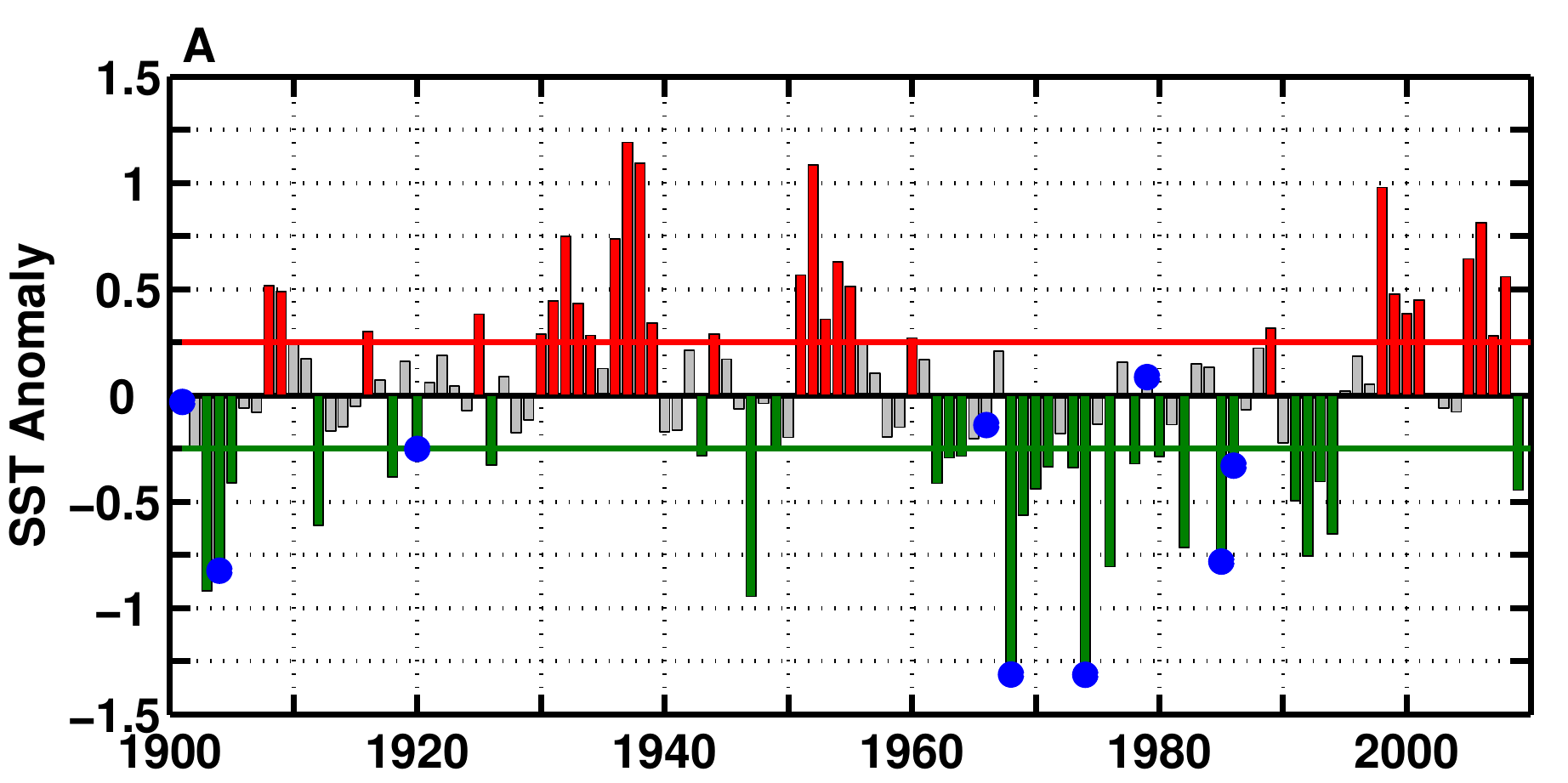}
  
\hspace*{0.05in}
\includegraphics[width=4.9in]{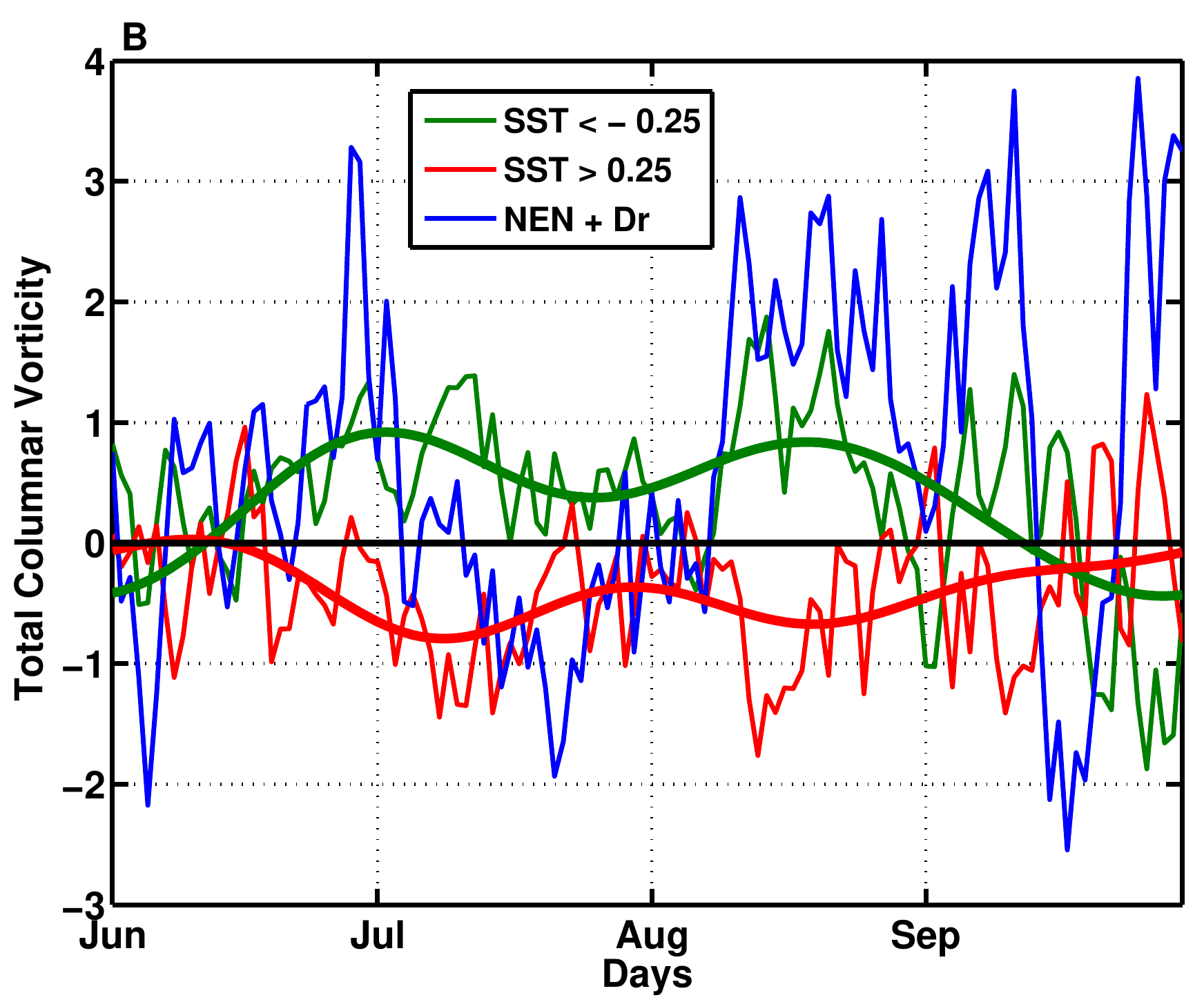}
\caption{(A) Long-term  (1900-2009) time series of  the North Atlantic
  SST anomalies  (averaged over  35N-55N; 310E-340E with  NEN droughts
  marked as  blue filled  circles.  The red  and green  bars represent
  ``warm''  and   ``cold''  years   when  the   SST  anomaly   is  $>$
  0.25$^\circ$C and  $< -$0.25$^\circ$C  ($\approx$ half  the standard
  deviation  of  long-term  SST  data), respectively.  Each  of  these
  categories comprise  of approximately 30  years. (B) Time  series of
  anomalies  of  daily total  columnar  (200,  300,  500 and  700  mb)
  vorticity  ($\times$ 10$^5$)  averaged over  this area  during JJAS,
  composited over the ``warm'' and  ``cold'' years.  The thick red and
  green curves  represent the respective first  three harmonics.  Time
  series of anomalies of daily total-columnar vorticity over the North
  Atlantic box for NEN + Dr years is shown in blue.}
\label{fig:fig_s9}
\end{figure}

\clearpage

\centerline{\large{\bf SUPPLEMENTARY TABLES}}


\begin{table}[hbpt]
  \begin{center}
    \caption{The list of Indian monsoon years where the seasonal (June
      through September; JJAS) rainfall anomaly  is less than -10\% of
      the long-term (1901-2015) mean, and  is classified as a drought.
      The anomalies  shown are  based on  the IITM  homogeneous Indian
      monthly  rainfall  dataset  \cite{iitmdataset}  for  the  period
      1901-2015.  The droughts that occurred during an El Ni\~{n}o (no
      El Ni\~{n}o) year are marked in red (blue) (see also Fig.~1c).}
    \medskip
  \begin{tabular}{||c|c||}\hline
    Year & Seasonal Rainfall Anomaly\\
    & (\% of long-term mean) \\\hline
\textcolor{black}{1901} & \textcolor{black}{-15}  \\\hline
\textcolor{black}{1904} & \textcolor{black}{-12}    \\\hline    
\textcolor{red}{1905} & \textcolor{red}{-16}  \\\hline  
\textcolor{red}{1911} & \textcolor{red}{-14}    \\\hline    
\textcolor{red}{1918} & \textcolor{red}{-24}    \\\hline    
\textcolor{black}{1920} & \textcolor{black}{-16}  \\\hline  
\textcolor{red}{1941} & \textcolor{red}{-15}  \\\hline  
\textcolor{red}{1951} & \textcolor{red}{-13}    \\\hline    
\textcolor{red}{1965} & \textcolor{red}{-17}    \\\hline    
\textcolor{black}{1966} & \textcolor{black}{-13}  \\\hline  
\textcolor{black}{1968} & \textcolor{black}{-11}  \\\hline  
\textcolor{red}{1972} & \textcolor{red}{-23}    \\\hline    
\textcolor{black}{1974} & \textcolor{black}{-12}  \\\hline  
\textcolor{black}{1979} & \textcolor{black}{-17}  \\\hline  
\textcolor{red}{1982} & \textcolor{red}{-14}    \\\hline    
\textcolor{black}{1985} & \textcolor{black}{-11}  \\\hline  
\textcolor{black}{1986} & \textcolor{black}{-13}  \\\hline  
\textcolor{red}{1987} & \textcolor{red}{-18}    \\\hline    
\textcolor{red}{2002} & \textcolor{red}{-22}    \\\hline    
\textcolor{red}{2004} & \textcolor{red}{-13}    \\\hline    
\textcolor{red}{2009} & \textcolor{red}{-22}    \\\hline    
\textcolor{black}{2014} & \textcolor{black}{-14}  \\\hline  
\textcolor{red}{2015} & \textcolor{red}{-14}    \\\hline    
  \end{tabular}
  \label{tab:tab_s1}
  \end{center}
  \end{table}

\clearpage

\begin{table}[hbpt]
  \centering
  \caption{Years when  the North  Atlantic SST (averaged  over 35-55N,
    310-340E)during JJAS was negative, but did not result in an Indian
    monsoon drought, and the corresponding seasonal rainfall anomaly.}
  \begin{tabular}{||c|c|c||}\hline
  Year & North Atlantic & Indian Monsoon \\
  &  SST Anomaly  & Seasonal Rainfall Anomaly\\
  &   ($^\circ$C)  & (\% of long-term mean) \\\hline
1903  &  -0.9   &     1\\\hline
1912  &  -0.6   &    -5.4\\\hline
1962  &  -0.4   &    -5\\\hline
1969  &  -0.6   &    -2.5\\\hline
1976  &  -0.8   &     0.5\\\hline
1991  &  -0.5   &    -7.9\\\hline
1992  &  -0.7   &    -7.9\\\hline
  \end{tabular}
  \label{tab:tab_s2}
\end{table}


\textcolor{black}{
\begin{landscape}
\begin{table}[hbpt]
  \centering
  \caption{Statistical  significance  of   anomalies  of  daily  total
    columnar    vorticity     during    9     NEN    +     Dr    years
    (\textcolor{red}{Fig. \ref{fig:fig_s9}b}; see also ``Supplementary
    Text'').  $\mu$  represents ``population''  mean, $\overline{V_N}$
    and $S_{V_N}$ represent the sample  mean and standard deviation of
    $N$ years  of daily  observations for the  period of  interest (in
    bold in  column 1)  during JJAS.   ``Cold'' years  represent those
    years when the North Atlantic  SST anomaly $< -0.25^{\circ}$C (see
    \textcolor{red}{Fig.  \ref{fig:fig_s9}a}).  All tests are based on
    a  $t$-distribution. The  second and  third tests  are based  on a
    two-sample $t$-test for comparing means \cite{ross2009}.}
  \begin{tabular}{||c|c|c|c|c||}\hline
Description &   Hypothesis & Sample Statistics     & Test Statistic & $p$-value \\
            &              &   & Value          & \\\hline
            &              &             &                & \\
{\bf Test1:} Comparison with  & $H_0$: ${\mu_{\omega}} = 0$ & $\overline{V_9}= 6.3 \times 10^{-6}$  &  & \\
long-term {\bf JJAS} mean   & $H_1$: ${\mu_{\omega}} \neq 0$& $S_{V_9} = 4\times 10^{-5}$    & 5.2 & ${\rm 2P\left(|t_{1097}| > 5.2\right)}$ \\
(2-sided)&            &       N = 9$\times$122 = 1098    &      & $\approx 0$  \\
            &              &             &                & \\\hline
            &              &             &                & \\
{\bf Test2:} Comparison with {\bf JJAS} &     & $\overline{V_9}= 6.3 \times 10^{-6}$            &        & \\
mean during ``cold'' & $H_0$: $\mu_{\omega}^{\rm
  NEN+Dr} = \mu^{\rm Cold}_{\omega}$ &   $\overline{V_{30}}= 3.7 \times 10^{-6}$  &   &\\
North Atlantic years  & $H_1$: $\mu_{\omega}^{\rm NEN+Dr} < \mu_{\omega}^{\rm Cold}$  & $S_{V_9} = 4\times 10^{-5}$    &  1.7 &
${\rm P\left(t_{N_9 + N_{30} -2} > 1.7\right)}$ \\
(1-sided)      &             &  $S_{V_{30}} = 4.5\times 10^{-5}$        &      & \\
               &             &  $N_9$ = 9 $\times$ 122 = 1098        &      & $\approx$ 4.6\%\\
               &             &  $N_{30}$ = 30 $\times$ 122 = 3660        &      & \\
&              &             &                & \\\hline
&              &             &                & \\
{\bf Test3:} Comparison with {\bf August}  &   &  $\overline{V_9}= 1.9 \times 10^{-5}$           &        & \\
{\bf episode mean (days 70-90)} during & $H_0$: $\mu_{\omega}^{\rm NEN+Dr} = \mu_{\omega}^{\rm Cold}$ &  $\overline{V_{30}}= 0.97 \times 10^{-5}$   &   &\\
``cold'' North Atlantic years  & $H_1$: $\mu_{\omega}^{\rm NEN+Dr} < \mu_{\omega}^{\rm Cold}$  & $S_{V_9} = 3.4\times 10^{-5}$     & 2.8  & ${\rm P\left(t_{N_9 + N_{30} -2} > 2.8\right)}$\\
(1-sided)      &             &    $S_{V_{30}} = 4\times 10^{-5}$    &         & \\
               &             &  $N_9$ = 9 $\times$ 21 = 189        &      & $\approx$ 0.3\%\\
               &             &  $N_{30}$ = 30 $\times$ 21 = 630        &      & \\
&              &             &                & \\\hline
  \end{tabular}
  \label{tab:tab_s3}
\end{table}
\end{landscape}
}





\begin{thebibliography}{10}
\bibitem{jjas_vort}
\bibinfo{title}{{The June and August episodes in NEN + Dr years (blue curve in
  Fig.~S9b) are strong enough to make the seasonal response (Fig. 3) seem like
  a deep tropospheric cyclonic circulation.}} .
\end{thebibliography}
\end{document}